\documentclass{emulateapj}
\usepackage{natbib}\usepackage{times}

\begin{document}

\shorttitle{The supernova rate does not match the star formation rate}
\shortauthors{Horiuchi et al.}
\title{The Cosmic Core-collapse Supernova Rate does not match the Massive-Star Formation Rate}
\author{Shunsaku Horiuchi\altaffilmark{1,2}, John F. Beacom\altaffilmark{1,2,3}, Christopher S.~Kochanek\altaffilmark{2,3}, Jose~L.~Prieto\altaffilmark{4,5}, K.~Z.~Stanek\altaffilmark{2,3}, \\ Todd A.~Thompson\altaffilmark{2,3,6}}
\altaffiltext{1}{Dept.\ of Physics, The Ohio State University, 191 W.\ Woodruff Ave., Columbus, OH 43210 \\URL: http://www.physics.ohio-state.edu/}
\altaffiltext{2}{Center for Cosmology and Astro-Particle Physics, The Ohio State University, 191 W.\ Woodruff Ave., Columbus, OH 43210 \\ URL: http://ccapp.osu.edu/}
\altaffiltext{3}{Dept.\ of Astronomy, The Ohio State University, 140 W.\ 18th Ave., Columbus, OH 43210 \\URL: http://www.astronomy.ohio-state.edu/}
\altaffiltext{4}{Carnegie Observatories, 813 Santa Barbara St., Pasadena, CA, 91101 \\ URL: http://obs.carnegiescience.edu/}
\altaffiltext{5}{Hubble and Carnegie-Princeton Fellow}
\altaffiltext{6}{Alfred P.~Sloan Fellow}

\email{horiuchi@mps.ohio-state.edu}

\begin{abstract}
We identify a ``supernova rate problem'': the measured cosmic core-collapse supernova rate is a factor of $\sim 2$ smaller (with significance $\sim 2 \sigma$) than that predicted from the measured cosmic massive-star formation rate. The comparison is critical for topics from galaxy evolution and enrichment to the abundance of neutron stars and black holes. We systematically explore possible resolutions. The accuracy and precision of the star formation rate data and conversion to the supernova rate are well supported, and proposed changes would have far-reaching consequences. The dominant effect is likely that many supernovae are missed because they are either optically dim (low-luminosity) or dark, whether intrinsically or due to obscuration. We investigate supernovae too dim to have been discovered in cosmic surveys by a detailed study of all supernova discoveries in the local volume. If possible supernova impostors are included, then dim supernovae are common enough by fraction to solve the supernova rate problem. If they are not included, then the rate of dark core collapses is likely substantial. Other alternatives are that there are surprising changes in our understanding of star formation or supernova rates, including that supernovae form differently in small galaxies than in normal galaxies. These possibilities can be distinguished by upcoming supernova surveys, star formation measurements, searches for disappearing massive stars, and measurements of supernova neutrinos. 
\end{abstract}


%%%%%%%%%%%%%%%%%%%%%%%%%%%%%%%%%%%%%%%%%%%%%%%%%%%%%%%%
%%%%%%%%%%%%%%%%%%%%%%%%%%%%%%%%%%%%%%%%%%%%%%%%%%%%%%%%
\section{Introduction}

Core-collapse supernovae (CC SNe) are extremely important to many areas of astrophysics. They are responsible for the majority of heavy elements \citep[e.g.,][]{1986A&A...154..279M}, are associated with dust production \citep[e.g.,][]{2001MNRAS.325..726T}, and could dominate winds and feedback in galaxy formation \citep[e.g.,][]{2006MNRAS.373..571F}. Observations of CC SNe and their progenitors test our understanding of stellar evolution (see, e.g., the progenitor-SN map of \citealp{2007ApJ...656..372G}), and the extreme densities and temperatures reached in CC SNe offer the opportunity to study the physics of weakly interacting particles, formation of compact objects, and related nuclear physics \citep[see, e.g.,][for recent reviews]{1990PhR...198....1R,2000PhR...333..593R,2006RPPh...69..971K,2007PhR...442...38J,2007PhR...442..109L}. However, there is much that is not yet understood.

%---------------------------------------------------------------
\begin{figure}[b]
\centering\includegraphics[width=\linewidth,clip=true]{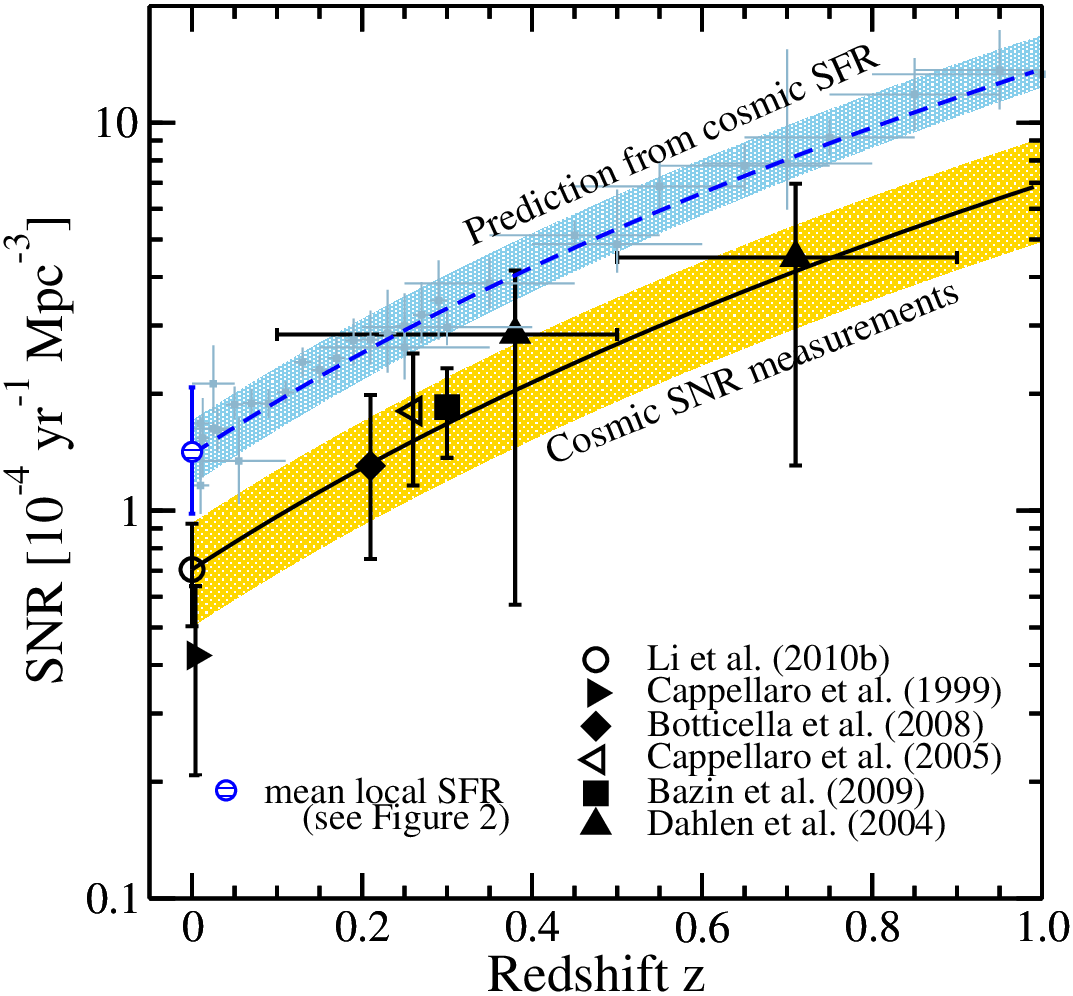}
\caption{Comoving SNR (all types of luminous core collapses including Type II and Type Ibc) as a function of redshift. The SNR predicted from the cosmic SFR fit and its supporting data \citep{2006ApJ...651..142H}, as well as that predicted from the mean of the local SFR measurements, are plotted and labeled. The fit to the measured cosmic SNR, with a fixed slope of $(1+z)^{3.4}$ taken from the cosmic SFR, is shown with the uncertainty band from the LOSS measurement. The predicted and measured cosmic SNR are consistently discrepant by a factor $\sim2$: the supernova rate problem. However, rates from SN catalogs in the very local volume do not show such a large discrepancy (see Figure \ref{fig:SNRate}).}\label{fig:comparison}
\end{figure}	
%---------------------------------------------------------------

One of the outstanding questions is the initial conditions corresponding to optically luminous, dim, and dark CC SNe \citep[see, e.g.,][]{2007PhR...442...38J}. It is expected that some CC SNe will be intrinsically dim or even dark in the optical bands. One possibility is the collapse of stars near the $8\rm{M_\odot}$ threshold via an electron capture trigger \citep{1980PASJ...32..303M,1984ApJ...277..791N,2008ApJ...675..614P}. Another possibility is the collapse proceeding directly to a black hole. Theoretical studies indicate that progenitors with $M \gtrsim 40 \, {\rm M_\odot}$ may promptly form black holes \citep{1999ApJ...522..413F}, with additional dependencies on the stellar metallicity and rotation of the progenitor core \citep{2003ApJ...591..288H}. CC SNe that are intrinsically luminous can also appear as optically dim CC SNe if they are heavily enshrouded by dust. Due to their darkness, these collapses are difficult to study directly. 

Precise measurements of the cosmic star formation rate (SFR) and the cosmic CC SN rate (SNR; we include all Type II and Type Ibc SNe) provide new information about CC SNe. The SFR and SNR encode the birth rate of stars and the death rate of massive stars, respectively. Since the massive stars that give rise to CC SNe have cosmologically short lifetimes $\sim 30 \, (M / 8 M_{\rm \odot})^{-2.5}$ Myr, the cosmic SNR is expected to follow the same evolutionary trend in redshift as the cosmic SFR. Second, the SFR data show that massive stars are clearly present, so they must be dying even if not all of these deaths are optically luminous. Because the measured SNR are sensitive to only optically luminous CC SNe, the normalization of the SNR with respect to the SFR provides information on the frequency of massive stars' fates as either optically luminous or dim (and dark) CC SNe\footnote{Typically, dim (low-luminosity) CC SNe are loosely defined as CC SNe falling at the lowest end of the CC SNe luminosity function; here, we define them as dimmer than $M = -15$ mag. The two definitions yield similar results; see Section \ref{sec:dim}.}.

Early attempts to compare the evolution and normalization of the cosmic SFR and cosmic SNR were inconclusive due to their large uncertainties. \cite{2004ApJ...613..189D} found reasonable agreement between their measured SNR at $z \approx 0.3$ and $z \approx 0.7$ and a compilation of SFR data by \cite{2004ApJ...600L.103G}. Using an improved compilation of modern cosmic SFR data, \cite{2006ApJ...651..142H} noted that the measured cosmic SNR \citep{1999A&A...351..459C,2004ApJ...613..189D,2005A&A...430...83C} were somewhat lower than those predicted from the SFR. Similar conclusions were reached by \cite{2007MNRAS.377.1229M} and \cite{2008A&A...479...49B}. 

In recent years, measurements of the cosmic SFR and cosmic SNR have rapidly improved. The cosmic SFR has been measured using multiple indicators by many competing groups. The accuracy and precision of the cosmic SFR has been documented \citep[e.g.,][]{2006ApJ...651..142H} and is supported by recent data \citep[e.g.,][]{2009ApJ...707.1740P,2010ApJ...718.1171R,2011ApJ...726..109L,2011arXiv1104.0929B}. The Lick Observatory Supernova Search (LOSS) has recently published the best measurement of the cosmic SNR at low redshifts, using CC SNe collected over many years of systematically surveying galaxies within $\sim 200$ Mpc \citep{2011MNRAS.412.1419L,2011MNRAS.412.1441L,2011MNRAS.412.1473L,2011MNRAS.tmp..307M}. The Supernova Legacy Survey (SNLS) has published the most precise SNR measurement at higher redshifts, using a large sample of CC SNe collected in their extensive rolling search of four deep fields \citep{2009A&A...499..653B}. 

Based on the latest data, it has become clear that the measured cosmic SFR and the measured cosmic SNR both increase by approximately an order of magnitude between redshift $0$ and $1$, confirming our expectation that the progenitors of CC SNe are short-lived massive stars \citep[e.g.,][]{2009A&A...499..653B,2011MNRAS.412.1473L}. On the other hand, the comparison of the normalizations of the latest SFR and SNR data has been left for future work. We perform this here for the first time. As illustrated in Figure \ref{fig:comparison}, the SNR predicted from the cosmic SFR is a factor of $\sim 2$ larger than the cosmic SNR measured by SN surveys; we term this normalization discrepancy the ``supernova rate problem.'' Both the predicted and measured SNR are of \emph{optically luminous} CC SNe, so the two can be directly compared. The lines in Figure \ref{fig:comparison} are fits to the SFR and SNR data, respectively\footnote{Technically, the SNR line shown is not a fit, but is a conservative estimate based on the SNR measurement of LOSS; see Section \ref{sec:SNRate}.}. The discrepancy persists over all redshifts where SNR measurements are available\footnote{However, in the local $\lesssim 25$ Mpc volume, the SNR derived from SN catalogs does not show such a large discrepancy, supporting earlier claims that the true cosmic SNR is as large as predicted \citep[e.g.,][]{2009PhRvD..79h3013H,2010ARNPS..60..439B}.}. 

The nominal uncertainties on the fits (shaded bands) are smaller than the normalization discrepancy, and the significance of the discrepancy is at the $\sim 2 \sigma$ level. At high redshift, where the uncertainties of the SNR measurements are largest, the statistical significance is weaker. However, it is remarkable how well the cosmic SNR measurements adhere to the expected cosmic trend---much better than their uncertainties would suggest. Indeed, the measurements of \cite{2004ApJ...613..189D} have been supported by recent unpublished results and with reduced uncertainties \citep{2010AAS...21543023D}. We therefore consider the fits to be a good representation, i.e., the supernova rate problem persists over a wide redshift range. We systematically examine resolutions to the supernova rate problem, exploring whether the cosmic SNR predicted from the cosmic SFR is too large, or whether the measurements underestimate the true cosmic SNR, or a combination of both. 

In Section 2, we describe the predicted and measured cosmic SNR in detail and substantiate the discrepancy. In Section 3, we discuss possible causes. In Section 4, we discuss our results and cautions. We summarize and discuss implications in Section 5. Throughout, we adopt the standard $\Lambda$CDM cosmology with $\Omega_m=0.3$, $\Omega_\Lambda=0.7$, and $H_0=73$ km s$^{-1}$ Mpc$^{-1}$. 

%%%%%%%%%%%%%%%%%%%%%%%%%%%%%%%%%%%%%%%%%%%%%%%%%%%%%%%%
%%%%%%%%%%%%%%%%%%%%%%%%%%%%%%%%%%%%%%%%%%%%%%%%%%%%%%%%
\section{Normalization of the Cosmic SNR} \label{sec:SNRate}

\begin{deluxetable*}{ccccccccl}
\tablecaption{Summary of Cosmic SNR Measurements \label{table:SNRate}}
\tablehead{ 
Redshift 	&  Rate ($10^{-4}$ $h_{73}^3$ yr$^{-1}$ Mpc$^{-3}$)	& $N_{\rm gal}$	& Cadence (days)	& $N_{\rm CCSN}$	& Host Extinction	& $m_{\rm lim}$ (mag)	& $M_{\rm lim}$ (mag) 	&  Reference }
\startdata
0		& $0.42 \pm 0.20$							& $\sim 10^4$	& -			& 67					& Y	& $m_R  \sim 16$	& - 				&  1\tablenotemark{a}	  	\\ \\
0		& $0.705^{+0.099}_{-0.097}(^{+0.164}_{-0.149})$	& 10121		& $\sim 9$	& 440				& N	& $m_R  \sim 19$	& $M_R \sim -15.0$	&  2	\\ \\
0.21		& $1.31^{+0.49}_{-0.37}(^{+0.48}_{-0.41})$		& 43283		& $\sim 120$	& 46.1\tablenotemark{b}	& Y	& $m_R  \sim 23$	& $M_R \sim -16.7$	&  3	\\ \\
0.26		& $1.81^{+0.91}_{-0.79}$						& 11300		& -			& 19.5				& N	& $m_V  \sim 23$	& $M_V \sim -17.1$	&  4\tablenotemark{a}	\\ \\
0.3		& $1.85^{+0.34}_{-0.34}(^{+0.34}_{-0.34})$		& Volumetric	& $\sim 7$	& 117 				& Y	& $m_i  \sim 24$ 	& $M_V \sim -16.4$	&  5	\\ \\
0.1--0.5	& $2.85^{+1.00}_{-0.85}(^{+0.85}_{-2.11})$		& Volumetric	& $\sim 45$	&  6					& Y	& $m_z  \sim 26$	& $M_R \sim -14.8$	&  6	\\  \\
0.5--0.9	& $4.49^{+1.17}_{-1.20}(^{+2.18}_{-2.95})$		& Volumetric	& $\sim 45$	& 10					& Y	& $m_z  \sim 26$	& $M_V \sim -16.0$	&  6	\\
\tableline
\tablenotetext{a}{We adopt the rate per $B$-band luminosity to volumetric rate conversion used by Bazin et al.~(2009).}
\tablenotetext{b}{Includes the CC SN sample of Cappellaro et al.~(2005).}
\tablecomments{Systematic error estimates are given in parentheses. The limiting absolute magnitudes are calculated from the limiting apparent magnitude at the distance of the data point. The scope of the rate measurements depend on a combination of limiting magnitude, SN luminosity function, and dust correction (see the text). }
\tablerefs{(1) \cite{1999A&A...351..459C}; (2) \cite{2011MNRAS.412.1473L}; (3) \cite{2008A&A...479...49B}; (4) \cite{2005A&A...430...83C}; (5) \cite{2009A&A...499..653B}; (6) \cite{2004ApJ...613..189D}.}
\end{deluxetable*}

The cosmic SNR is calculated from the cosmic SFR using knowledge of the efficiency of forming CC SNe. The most recent SFR is traced by the most massive stars that have the shortest lifetimes. The primary indicators of massive stars---H$\alpha$, UV, FIR, and radio---are routinely used, with dust corrections where necessary, to study the populations of massive stars. However, since the \emph{total} SFR is dominated by stars with smaller masses, the SFR derived from massive stars must be scaled upward according to the initial mass function (IMF); for example, for a given massive stellar population, an IMF that is more steeply falling with mass will yield a larger total SFR compared to a shallower IMF. The scaling is done with the use of calibration factors derived from stellar population synthesis codes that calculate the radiative output from a population of stars following an assumed IMF \citep[see, e.g.,][]{1998ARA&A..36..189K}. 

We adopt the dust-corrected SFR compilation of \cite{2006ApJ...651..142H}. Their data are well fit by a smoothed broken power law of the form \citep{2008ApJ...683L...5Y}
%-------------
\begin{eqnarray} \label{eq:fit}
\dot{\rho}_*(z)
& = &  \dot\rho_0 \left[(1 + z)^{{a}{\eta} } + \left(\frac{1 + z}{B}\right)^{{b}{\eta}}
+ \left(\frac{1 + z}{C}\right)^{{c}{\eta} } \, \right]^{1/\eta},
\end{eqnarray}
%-------------
where $B = (1 + z_1)^{1-a/b}$, $C =(1 + z_1)^{(b-a)/c} (1 + z_2)^{1-b/c}$. We adopt $\dot{\rho}_0=0.016\, h_{73} \, {\rm M_\odot \, Mpc^{-3} \, yr^{-1}} $ for the cosmic SFR at $z=0$, as well as the parametrization $a=3.4$, $b=-0.3$, $c=-3.5$, $z_1=1$, $z_2=4$, and $\eta=-10$. These choices are applicable for the Salpeter A IMF, which is a modified Salpeter IMF with a turnover below $1 {\rm M_\odot}$ \citep{2003ApJ...593..258B}. The scaling from a Salpeter IMF is $\approx 0.77$. The $1 \sigma$ uncertainty on $\dot{\rho}_0$ is approximately $\pm 25$\% \citep{2006ApJ...651..142H}. The fit is in good agreement with a range of recent SFR measurements over a range of redshifts, for example using H$\alpha$ \citep{2011ApJ...726..109L}, UV \citep{2007ApJS..173..267S}, IR \citep{2009ApJ...707.1740P,2010ApJ...718.1171R}, and X-ray \citep{2009ApJ...696.2206W}. Most recently, a thorough study of the local SFR at $z \approx 0.05$ has been performed by \cite{2011arXiv1104.0929B} using combined observations in UV and IR. Adjusting to our chosen cosmology and IMF, the derived SFR is $ 0.0193 \pm 0.0012 \, h_{73} \, {\rm M_\odot \, Mpc^{-3} \, yr^{-1}} $, in good agreement with our parametrization. At $\pm 6$\%, the uncertainty is a great improvement over many previous measurements and our adopted uncertainty on $\dot{\rho}_0$.
 
The comoving volumetric SNR is determined by multiplying Equation (\ref{eq:fit}) by the efficiency of forming CC SNe. This is the number of stars that eventually explode as CC SNe per unit stellar mass formed in a burst of star formation. It is largely governed by the mass range for CC SNe, from $M_{\rm min}$ to $M_{\rm max}$, resulting in a SNR of
%-------------
\begin{equation} \label{eq:SNRate}
R_{\rm SN}(z) = \dot{\rho}_*(z)
\frac{\int_{M_{\rm min}}^{M_{\rm max}}\psi(M)dM}{\int_{0.1}^{100} M \psi(M)dM},
\end{equation}
%-------------
where $\psi(M)$ is the IMF, defined over the main-sequence mass range 0.1--100 $M_{\odot}$. The IMF is defined such that $\psi(M) dM$ gives the number of stars in the mass range $M$ to $M+dM$. Due to the steeply falling nature of the IMF, the lower mass limit $M_{\rm min}$ is the most important parameter. Note that this selection of the relevant mass range is effectively the inverse process of the scaling from the massive-star SFR to the total SFR. In fact, the stellar mass range probed by the SFR indicators is comparable to the mass range giving rise to CC SNe. Thus, variations in the IMF should have only a small effect on the predicted SNR. 

The predicted cosmic SNR is shown in Figure \ref{fig:comparison}. We assumed canonical parameters for optically luminous CC SNe, $M_{\rm min} = 8 M_{\odot}$ and $M_{\rm max}= 40 M_{\odot}$. The SFR to SNR conversion coefficient is then $0.0088/{\rm M_\odot}$, yielding $R_{\rm SN} (z=0) \approx 1.4 \times 10^{-4} \, {\rm yr^{-1} \, Mpc^{-3}} $. The uncertainty band shown is the 1$\sigma$ uncertainty in the SFR fit of \cite{2006ApJ...651..142H}. The actual data of the compilation is also shown, similarly converted to a SNR: the dust-corrected UV from Sloan Digital Sky Survey \citep[SDSS;][]{2005MNRAS.358..441B}, Galaxy Evolution Explorer \citep[GALEX;][]{2005ApJ...619L..43A,2005ApJ...619L..47S}, COMBO17 \citep{2003A&A...401...73W}, and H$\alpha$-derived measurement \citep{2006ApJ...649..150H}. The mean of the local SFR measurements (Section \ref{sec:localSFR}) is also converted and shown.

There are two main approaches for collecting CC SNe and measuring the cosmic SNR. In the first, the same patch of sky is periodically observed, locating CC SNe within a volume, limited only by flux. In the second, a pre-selected sample of galaxies is periodically observed, and the rate per unit galaxy size (mass or light) is converted to a volumetric rate using the galaxy mass or luminosity density. In the former, the completeness of the CC SN sample is readily definable and the volumetric rate is derived directly, but it requires a sufficiently wide or deep field to collect CC SN statistics. In the latter, there is a bias against CC SNe in small galaxies, but, for local surveys, it maximizes the CC SN discovery rate given the closer target volume. In Table \ref{table:SNRate}, we summarize measurements of the cosmic SNR in the literature, showing the rate, the number of galaxies sampled (where appropriate), the cadence, the number of CC SNe used for the rate analysis, whether host galaxy extinction corrections are made, and the limiting magnitudes of the surveys; the survey characteristics vary considerably. 

The limiting magnitude of SN surveys is typically defined as the magnitude at which the detection efficiency is $50$\% \citep[e.g.,][]{2004ApJ...613..189D,2009A&A...499..653B,2011MNRAS.412.1473L}. For \cite{2008A&A...479...49B}, we quote the limiting magnitude for a seeing of $1$ arcsec, their average. The limiting absolute magnitude is estimated from the limiting apparent magnitude and the distance of the data point (for \citealp{2004ApJ...613..189D}, we use the redshift that divides the distance bin in two equal volumes). This estimate is only indicative, because dimmer CC SNe can be discovered at smaller distances. Furthermore, SNR measurements adopt a CC SNe luminosity function, usually but not always derived from the SN survey data itself, to correct for missing dimmer CC SNe. The SNR measurements should therefore be treated as measurements of CC SNe within the luminosity limits of the SN luminosity function adopted. We show apparent magnitudes in the observed-frame band and the absolute magnitudes in the supernova rest-frame band. For LOSS, we quote the limiting magnitude of their volume-limited subsample and not their full sample, since the former is used to correct the latter. 

The most reliable SNR measurement is that by LOSS \citep{2011MNRAS.412.1419L,2011MNRAS.412.1441L,2011MNRAS.412.1473L,2011MNRAS.tmp..307M}: they have excellent cadence and the largest published SN sample. They use a large subsample of 101 volume-limited CC SNe to construct a modern SN luminosity function. It is complete to $M_R \sim -15$ mag, although it is corrected to go dimmer \citep[the LOSS-LF sample;][]{2011MNRAS.412.1441L}. Their uncertainty of approximately $\pm 30$\% is dominated by systematics, and originates mainly from limited knowledge of the $z\approx 0$ galaxy luminosity density needed to derive the volumetric rates from their rate per galaxy luminosity \citep{2011MNRAS.412.1473L}. Not surprisingly, this is comparable to the uncertainty on the cosmic SFR normalization at $z\approx 0$ \citep{2006ApJ...651..142H}. At higher redshifts, the most precise measurement is that of SNLS, which is a volume monitoring search with excellent cadence and CC SNe statistics \citep{2009A&A...499..653B}. The luminosity function of SNLS is in good agreement with that of LOSS-LF, although LOSS goes approximately $1.5$ mag fainter on the supernova luminosity function \citep{Rich2011}. Their result has been corrected for intrinsic extinction due to host inclination.

Measurements of the cosmic SNR are shown in Figure \ref{fig:comparison}. Those with host galaxy extinction corrections are indicated by filled symbols \citep{1999A&A...351..459C,2004ApJ...613..189D,2008A&A...479...49B,2009A&A...499..653B}, while those without are indicated by empty symbols \citep{2005A&A...430...83C,2011MNRAS.412.1473L}. The error bars combine the statistical and systematic uncertainties in quadrature. Unpublished results from \cite{2010AAS...21543023D} support their previous measurements \citep{2004ApJ...613..189D}. The high redshift measurements are further supported by recent results from the Subaru Deep Field which lie very close although with somewhat larger error bars \citep{2011arXiv1102.0005G}. Preliminary results from SDSS on the SNR at $z \lesssim 0.2$ also lie on the trend of the other rate measurements \citep{2009AAS...21342502T}.  

A fit to the SNR measurements, excluding the measurement by \cite{1999A&A...351..459C} and assuming a slope identical to the cosmic SFR, yields a best-fit normalization $0.70 \times 10^{-4} \, {\rm yr^{-1} \, Mpc^{-3}}$ with $\chi^2_{\rm min} \approx 0.5$ for 5 degrees of freedom. The normalization is necessarily low due to the fact that some data have not been corrected for host-extinction. The $\chi^2$ is somewhat low, reflecting the nature of the large systematic uncertainties. The $1 \sigma$ confidence region corresponding to $\chi^2 < \chi^2_{\rm min} + \Delta \chi^2$, where $\Delta \chi^2 = 1.00$ for a 1 parameter fit \citep[e.g.,][]{1976ApJ...210..642A}, is smaller than the systematic uncertainty of LOSS. To remain conservative, we show the combined statistical and systematic uncertainty of LOSS and not of the fit. As a check, we also let the slope vary and find comparable results: a best-fit normalization of $0.72 \times 10^{-4} \, {\rm yr^{-1} \, Mpc^{-3}}$ and a slope of $3.3$, with $\chi^2_{\rm min} \approx 0.5$ for 4 degrees of freedom. 

The comparison of the predicted and measured cosmic SNR, and the relative sizes of the uncertainties, demonstrate two key points: they evolve similarly in redshift, and there is a systematic normalization mismatch. 

%%%%%%%%%%%%%%%%%%%%%%%%%%%%%%%%%%%%%%%%%%%%%%%%%%%%%%%%
%%%%%%%%%%%%%%%%%%%%%%%%%%%%%%%%%%%%%%%%%%%%%%%%%%%%%%%%
\section{Possible Explanations}\label{sec:uncertainty}

%======================================================================================
\subsection{Is the Cosmic SFR too High?}\label{sec:sfr}

We address the scatter in and check the normalization of the cosmic SFR of \cite{2006ApJ...651..142H}. In particular, the scatter mainly originates from the different indicators used to derive the SFR, but also the intrinsic uncertainty in the calibration of each indicator. The normalization is strongly influenced by the choice of IMF as well as dust corrections. Below, we discuss these in turn.

The SFR compilation includes measurements from H$\alpha$, UV, FIR, and radio indicators. The scatter in the SFR fit is dominated by the scatter in the measurements from these different groups; it is typically $\pm 30$\% in the redshift range $0 \lesssim z \lesssim 1$ \citep{2006ApJ...651..142H}. At the lowest redshift, the scatter approaches the intrinsic uncertainties in the calibrations of each indicators. For the bulk of the galaxies the continuous star formation approximation (i.e., star formation remains constant on time scales longer than the lifetimes of the dominant UV-emitting massive stars, $\sim 60$ Myr; \citealp{1998ARA&A..36..189K}) holds, so that the calibration uncertainties are $\pm 10$\%--$20$\%. If the assumption is incorrectly applied, as would be the case for young starburst galaxies, the SFR would be underestimated by $\sim 30$\%. Therefore, the scatter in the SFR derived from a single indicator and the scatter among different indicators are generally at the tens of percent level. 

The choice of IMF plays an important role in the normalization of the SFR, because the IMF is used to scale the observed massive-SFR to the total SFR. Adopting instead of the Salpeter A IMF a classic Salpeter IMF or a flatter Baldry-Glazebrook IMF, the total SFR normalization varies at the $\pm 30$\% level  \citep[e.g.,][]{2006ApJ...651..142H}. However, it is important to stress that this does not affect the predicted SNR normalization anywhere near as strongly, because the massive stars that are used to make the SFR measurements are close in mass to the progenitors of CC SNe. In other words, an increase or decrease of the SFR due to steeper or shallower IMF is nearly completely canceled by a respective decrease or increase in the fraction of massive stars. The difference between the predicted SNR using the Salpeter A and the Baldry-Glazebrook IMF is less than $10$\% \citep{2009PhRvD..79h3013H}.

While extinction by dust raises the possibility for the SFR to be over-corrected for dust, it is not realistically possible that the SFR in the most important redshift range ($0 < z < 1$) is over estimated by a factor of $\sim 2$. Since interstellar dust absorbs UV emission and re-emits in the FIR, the UV-derived SFR must be corrected. In our adopted SFR compilation, the UV-derived SFR are dust-corrected by adding the FIR-derived SFR of \cite{2005ApJ...632..169L} at their respective redshifts. The accuracy of this extinction correction has been discussed using individual galaxies \citep{2006ApJS..164...38I} and by comparisons with independent H$\alpha$-derived SFR \citep{2003ApJ...586..794B}; see also \cite{2004ApJ...615..209H} and \cite{2006ApJ...651..142H}. The SFR corrected for dust in this way are consistent with recent SFR measurements in many wavebands and dust-correction methods (see Section \ref{sec:SNRate}). In the range $z \lesssim 0.4$, where the SNR normalization discrepancy is most significant, dust correction typically increases the UV-derived SFR by a factor of $\sim 2$. Thus, to explain the supernova rate problem by lowering the SFR, one would require almost no dust at these distances, which is inconsistent with our understanding of the FIR universe \citep{2001ARA&A..39..249H,2005ARA&A..43..727L}.

At higher redshifts ($0.5 \lesssim z \lesssim 1$) the dust correction to the UV data become larger. By $z \approx 1$, more than 80\% of the total SFR comes from the FIR. Eventually, the dust opacity in galaxies become high enough that the FIR luminosity itself measures the bolometric luminosity of the star formation activity. To reduce the SFR by a factor of $\sim 2$ under these conditions requires one to either have severe contamination of the FIR emission from non-star-forming sources or that the FIR to SFR calibration is incorrect. While it is known that there is a ``cirrus'' component in the FIR that is associated with more extended dust heated by a population of older stars, their relative contribution varies substantially from galaxy to galaxy. For spirals it can be as high as $50$\%--70\%, but for the star forming galaxies that dominate at high redshifts it is expected to be much smaller \citep{1987ApJ...314..513L,2002AJ....124.3135K}. The calibration uncertainties are $\pm10$\%--$20$\% as described above. 

Finally, the integrated SFR normalization can be checked against observable quantities \citep{1998ApJ...498..106M}, for example the stellar mass density and the extragalactic background light (EBL). The stellar mass density has been measured up to redshifts of a few and are typically a factor of $\sim 2$ smaller than those predicted from the SFR assuming the Salpeter A IMF \citep[see, e.g.,][]{2006ApJ...651..142H,2008MNRAS.385..687W}. While a potential explanation of this discrepancy is to reduce the cosmic SFR normalization by a factor of $\sim 2$, i.e., at the same time also solving the supernova rate problem, this is unlikely to be the correct explanation. First, the stellar mass density is dominated by low-mass stars rather than the high-mass stars constrained by the SFR, leaving the comparison to reveal more about the IMF slope rather than the SFR normalization. For example, an IMF with a high-mass slope shallower ($-2.15$; equivalent to the Baldry-Glazebrook IMF) than the Salpeter slope ($-2.35$) reconciles the discrepancy at $z < 0.4$ \citep{2008MNRAS.391..363W}. Second, it has been argued that observations could be underestimating the stellar mass density \citep[e.g.,][]{2004ApJ...610...45N,2001MNRAS.320..504S}. Third, as we discuss below, such an explanation creates a tension in explaining the EBL. 

The EBL is powered by moderately high mass stars (half of the stellar EBL is powered by stars with masses $M \gtrsim 3 M_\odot$), so that the IMF dependency is modest. Approximately 70\% of the EBL arises from the $z<1$ range we are most interested in, with only a few percent contribution from non-nucleosynthesis energy sources such as active galactic nuclei \citep[AGNs;][]{2006ApJS..163....1H}. This makes the EBL a useful probe of the integrated ($0 < z < 1$) SFR normalization. The minimum EBL has been derived indirectly by counting visible galaxies, while the maximum EBL has been directly measured \citep[see, e.g.,][]{2001ARA&A..39..249H}. We summarize the findings of \cite{2009PhRvD..79h3013H} here. Integrating from the FIR to UV bands yields a total EBL in the range $50$--$100 \, {\rm nW \, m^{-2} \, sr^{-1}}$, comparable with other previously evaluated estimates \citep[e.g.,][]{2007MNRAS.379..985F}. The EBL has recently been constrained by the observations of gamma rays from distant blazars to be between these extremes \citep{2006Natur.440.1018A,2008Sci...320.5884A,2010ApJ...723.1082A,2011ApJ...733...77O}. We adopt a nominal total EBL of $73 \, {\rm nW \, m^{-2} \, sr^{-1}}$ that respects the gamma-ray constraints. If secondary gamma rays produced en route by beamed cosmic rays are important \citep{2010APh....33...81E,2011ApJ...731...51E}, the EBL could be even higher. In comparison, the result calculated from the SFR of \cite{2006ApJ...651..142H}, using the PEGASE.2 population synthesis code \citep{1997A&A...326..950F}, is $88^{+36}_{-28} \, {\rm nW \, m^{-2} \, sr^{-1}}$ for the Salpeter A IMF and $78^{+31}_{-24} \, {\rm nW \, m^{-2} \, sr^{-1}}$ for the Baldry-Glazebrook IMF. Therefore, the predicted and measured total EBLs do not allow a factor of $\sim 2$ reduction of the SFR.

In summary, the uncertainties associated with SFR measurements are generally at the tens of percent level. The integrated cosmic SFR normalization has been cross-checked with the EBL at a similar precision, and does not allow the cosmic SFR to be decreased enough to explain the supernova rate problem.

%======================================================================================
\subsection{Does the Local SFR Differ from the Cosmic SFR?}\label{sec:localSFR}

The local ($\lesssim 100$ Mpc) SFR sets the birth rate of stars nearby and has important implications for our study of the local SNR. Here, we discuss whether the local SFR measurements are different from the cosmic SFR. Since the SFR are densities, they should not depend on distance, except for evolution with redshift, which is modest within $100$ Mpc. We address the validity of the comparison in light of expected cosmic variance and bias due to missing massive galaxies in small volumes, before discussing the results.

Cosmic variance falls approximately as distance to the power $-2$ \citep{1994MNRAS.267.1020P} and is minimal by several tens of Mpc. Normalizing the density power spectrum to the seven-year WMAP result of $\sigma_8 = 0.809$ \citep{2011ApJS..192...14J}, the density fluctuations within $30$ Mpc should only be $\sim 0.1$ of the mean density. Variations in smaller volumes are more likely. For example, it is not surprising that the galaxy $B$-band luminosity density in the $8$ Mpc volume is a factor of $1.7$--$2.0$ times the global luminosity density \citep[as derived from the SDSS and the Millennium Galaxy Catalog;][]{2004AJ....127.2031K}. 

Given the smaller volume, the rarest and most massive galaxies would be missing. However, the effect of missing their star formation is minimal. While massive galaxies have higher SFR, massive galaxies are more rare and have lower star formation per mass, so that the contribution to the total SFR peaks at galaxy mass $\sim 6 \times 10^{10} \, {M_\odot}$. Consider the volume within $40$ Mpc. Integrating the SDSS galaxy stellar mass function, the number of galaxies more massive than $4 \times 10^{11} \, {\rm M_\odot}$ is expected to be $\lesssim 1$. The fraction of the SFR missed by excluding galaxies above $4 \times 10^{11} \, {\rm M_\odot}$ is less than $10$\% \citep{2004MNRAS.351.1151B}. Therefore, the volume within several tens of Mpc is a fair representation of the global SFR.

\begin{figure}[tb]
\centering\includegraphics[width=\linewidth,clip=true]{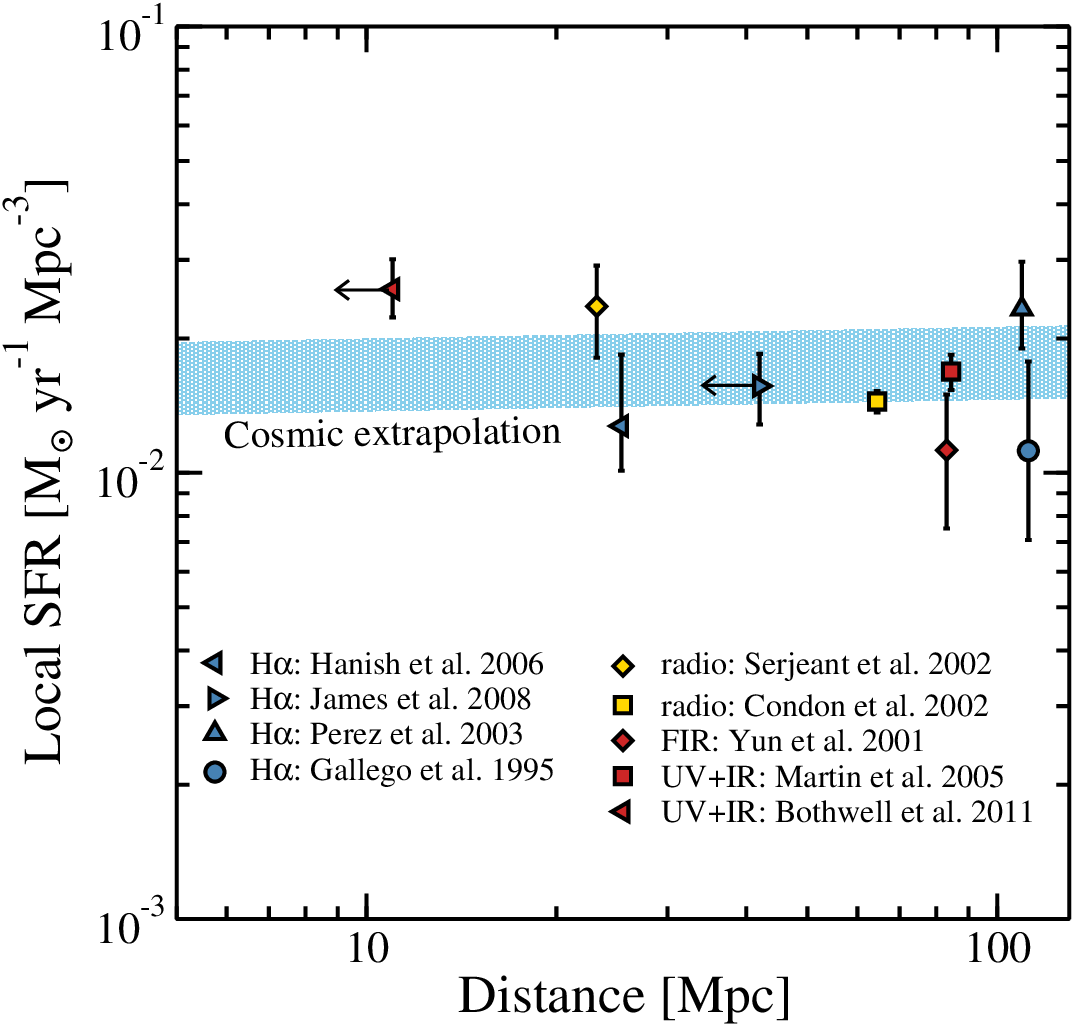}
\caption{Local SFR density as a function of distance. Measurements are shown at the mean distances of their respective galaxy samples, except for 11HUGS \citep{2011arXiv1104.0929B} and H$\alpha$GS \citep{2008A&A...482..507J} which are measurements within fixed distances and are shown at their distance limits (horizontal arrows). The SFR within $11$ Mpc is enhanced with respect to the cosmic SFR (shaded band) due to cosmic variance. Beyond $20$ Mpc, the measurements and their mean, $0.017^{+0.007}_{-0.005} \, {\rm M_\odot \, Mpc^{-3} \, yr^{-1}} $, are compatible with the cosmic SFR, and show no significant cosmic variance in this distance range. Shown for a Salpeter A IMF.} \label{fig:SFrate}
\end{figure}

In Figure \ref{fig:SFrate}, we show SFR measurements in the local volume, derived from H$\alpha$ \citep{1995ApJ...455L...1G,2003ApJ...591..827P,2006ApJ...649..150H,2008A&A...482..507J}, FIR \citep{2001ApJ...554..803Y}, radio \citep{2002MNRAS.330..621S,2002AJ....124..675C}, and combination of UV and IR \citep{2005ApJ...619L..59M,2011arXiv1104.0929B}. The measurements are corrected for dust and have been scaled to our chosen cosmology and IMF. At distances just beyond those shown, a recent thorough study using a combined UV and IR dataset find $ 0.0193 \pm 0.0012 \, h_{73} \, {\rm M_\odot \, Mpc^{-3} \, yr^{-1}} $ (at $z\approx 0.05$), in excellent agreement with the cosmic SFR and with small uncertainty \citep{2011arXiv1104.0929B}.

%---------------------------------------------------------------
%---------------------------------------------------------------

The SFR within $11$ Mpc has been measured by the $11$ Mpc H$\alpha$ and Ultraviolet Galaxy Survey \citep[$11$HUGS;][]{2008ApJS..178..247K,2011arXiv1104.0929B}. The value, $ 0.0262^{+0.0044}_{-0.0035} \, h_{73} \, {\rm M_\odot \, Mpc^{-3} \, yr^{-1}} $, shows an overdensity with respect to the cosmic SFR. The degree of overdensity is approximately a factor of $1.7$ and is within expectations of cosmic variance. It is smaller than but similar to the enhancement seen in the galaxy $B$-band luminosity density. The remaining measurements at larger distances are compatible within errors with the cosmic SFR. The mean of the measurements, omitting the $11$HUGS measurement because of the known overdensity, is $0.017^{+0.007}_{-0.005} \, {\rm M_\odot \, Mpc^{-3} \, yr^{-1}} $. This is in good agreement with the cosmic SFR. The uncertainty range has been conservatively taken to be the range of the  SFR measurements, because some of the measurements are in disagreement with each other, suggesting the true uncertainties are larger than reported for some individual measurements. We conclude that the local SFR data do not support large deviations from the cosmic SFR. 

In summary, SFR measurements in the local volume are in agreement within uncertainties with the better-measured cosmic SFR, except for the volume within $\lesssim 10$ Mpc which has a factor of $\lesssim 2$ enhancement. These results provide the important basis on which we interpret the local SNR in Section \ref{sec:dim}.

%======================================================================================
\subsection{Is the Fraction of Stars resulting in CC SNe too High?}\label{sec:massrange}

The predicted SNR depends on the number of stars that become CC SNe per unit stellar mass formed. To a first approximation, this depends on the mass range of stars that make CC SNe. There is observational evidence that a wide range of massive stars yield optically luminous CC SNe. Conceptually, the mass range is controlled by two parameters, the lower mass limit that is the boundary between the formation of a white dwarf and a neutron star, and the upper mass limit that is less well defined but could be the boundary between forming a neutron star and a stellar mass black hole. This is only approximate, since stellar evolution and SN simulations indicate that parameters such as rotation and metallicity also affect the outcome of core collapse, and prompt black hole formation may be accompanied by some transient phenomena, even if dim. For example, a rare group of highly rotating stars that collapse to black holes form the current paradigm for gamma-ray burst central engines \citep{1993ApJ...405..273W}. We assume that parameters other than mass are averaged out in integrated populations and that the mass is the primary parameter for the CC SNe outcome. The existence of CC SNe progenitors of a given mass does not require or prove that all progenitors of that mass produce similar CC SNe. 

The lower mass limit is newly supported by direct observations of CC SNe progenitors. Using archival images, it has been possible to identify the stellar progenitors of some nearby CC SNe \citep[see, e.g.,][for a review]{2009ARA&A..47...63S}. Using the stars' luminosities and other information, the stellar masses can be measured or limited. In \cite{2009MNRAS.395.1409S}, the authors reviewed 20 progenitors of Type IIP SNe and statistically found the progenitor mass limit $M_{\rm min}^{\rm IIP} = 8.5^{+1}_{-1.5} \, {\rm M_\odot}$. This is consistent with the highest masses estimated for white dwarf progenitors, $\sim 7 \, {\rm M_\odot}$ \citep{2008ApJ...676..594K,2009ApJ...693..355W}. Thus two different approaches seem to be converging to $M_{\rm min} \approx 8 \pm 1 \, {\rm M_\odot}$. This uncertainty affects the predicted SNR at the $\pm 20$\% level. 

On the other hand, the upper mass limit is less certain. Fortunately, as long as the upper limit is large, it does not strongly affect the predicted SNR. For example, to fully explain the normalization discrepancy in Figure \ref{fig:comparison} would require $M_{\rm max} \approx 13 \, {\rm M_\odot}$ for $M_{\rm min} = 8 \, {\rm M_\odot}$. Theoretically, the upper mass limit is $M_{\rm max} \approx 40 \, {\rm M_\odot}$ for a sub-solar metallicity star \citep{1999ApJ...522..413F}. Higher metallicity generates stronger mass loss prior to collapse and predicts the formation of a Wolf-Rayet (WR) star, which can extend $M_{\rm max}$ to $100 \, {\rm M_\odot}$ \citep{2003ApJ...591..288H}. The upper mass limit derived from the analysis of 20 Type IIP SN progenitors is $M_{\rm max}^{\rm IIP} = 16.5 \pm 1.5 \, {\rm M_\odot}$ \citep{2009MNRAS.395.1409S}. No progenitors of Type Ibc SNe have been directly identified, but the spatial distribution of CC SNe shows that Type Ibc progenitors must be more massive than those of Type IIP SNe \citep{2008MNRAS.390.1527A,2009MNRAS.399..559A}. It is widely expected that Type Ibc originate from evolved massive WR stars that have shed their envelopes. From Galactic stellar clusters and stellar clusters in the LMC, such WR stars are estimated to have evolved from main-sequence masses 25 ${\rm M_\odot}$ and above \citep{2000AJ....119.2214M,2001AJ....121.1050M}. Very luminous Type IIn have been observed to arise from the core-collapse of very massive stars of the luminous blue variable (LBV) type \citep{2007ApJ...656..372G,2009Natur.458..865G,2011ApJ...732...63S}, whose masses are $\gtrsim 20 {\rm M_\odot}$. Recent studies of the peculiar SN~1961V support its nature as a true CC SNe, implying its progenitor, a massive $\gtrsim 80 \, {\rm M_\odot}$ star, produced an optically luminous CC SNe \citep{2010arXiv1010.3704K,2010arXiv1010.3718S}. Therefore, there is substantial evidence for the upper mass limit being high.

However, the outcomes of the mass range from 17 to 25--30 $\rm M_\odot$ remain uncertain. \cite{2009MNRAS.395.1409S} noted that stars in this mass range are not found as CC SNe progenitors, and termed this the ``red supergiant problem.'' Future progenitor studies may discover CC SNe progenitors in this mass range. For example, the progenitor of the recent Type IIL SN~2009kr has been suggested to be $\sim 20 {\rm  M_\odot}$ (\citealp{2010ApJ...714L.254E}, although \citealp{2010ApJ...714L.280F} find $\sim 15 {\rm  M_\odot}$). It is also possible that the majority of these stars form black holes at core collapse and lead to dim or dark CC SNe \citep{2009MNRAS.395.1409S}. In Section \ref{sec:summary} we discuss how monitoring for the disappearance of high mass stars will allow observation of optically dark core collapses \citep{2008ApJ...684.1336K}. 

In summary, we adopt the nominal mass range for optically luminous CC SNe of $8$--$40 \, {\rm M_\odot}$, based on stellar and supernova simulations; the maximum mass range $8$--$100 \,{\rm M_\odot}$; and the conservative minimum mass range the combination of $8.5$--$16.5 \, {\rm M_\odot}$ and $25$--$40 \, {\rm M_\odot}$. Compared to the nominal  $8$--$40 \, {\rm M_\odot}$, the maximum and minimum mass ranges affect the predicted cosmic SNR by $+10$\% and $-30$\%, which are insufficient to explain the supernova rate problem. 

%======================================================================================
\subsection{Are Measurements Missing Luminous CC SNe?}\label{sec:luminous}

Luminous CC SNe (we henceforth define luminous CC SNe as CC SNe more luminous than $-15$ mag) that fall within the sensitivity of SN surveys can still be missed. Here, we discuss the potential impact of these missed luminous CC SNe.

Flux-limited surveys targeting a pre-selected sample of galaxies are naturally biased against small galaxies, and inevitably result in some CC SNe being missed. Even the LOSS survey views only $45$\% of the total stellar light accessible from the survey position \citep{2011MNRAS.412.1419L}. In the early works of \cite{1999A&A...351..459C} and updates by \cite{2005A&A...430...83C} and \cite{2008A&A...479...49B}, this has been corrected for by assuming that the CC SN rate per galaxy size (mass or luminosity) remains constant in galaxies having the same Hubble type or $B-K$ color. Therefore, the volumetric SNR could be determined even if the faint end of the galaxy luminosity function is not fully sampled, by measuring the CC SN rate in large galaxies and multiplying by an independently measured galaxy luminosity density that has been integrated to include small galaxies. However, more careful sampling by LOSS helped reveal the rate-size relation, where the SN frequency is not linearly proportional to the galaxy size but rather size to the power of $0.4$--$0.6$, the exact number depending on the SN Type and galaxy Hubble type \citep{2011MNRAS.412.1473L}. The rate-size relation is comparable to the trend of specific-SFR with galaxy mass \citep{2007ApJS..173..267S,2007ApJS..173..315S} and implies more emphasis on small galaxies.

However, the rate-size relation has been measured with galaxies down to $L_K \approx 2 \times 10^{10} \, {L_\odot}$ ($M_K \approx -22.4$ mag, approximately $1.3$ mag fainter than the Milky Way), and additional effects in smaller galaxies may not addressed. In particular, there is mounting evidence that CC SNe in small galaxies are unlike those in large normal galaxies. Some types of rare CC SNe occur only in low-metallicity dwarf galaxies \citep{2003A&A...400..499L,2006Natur.441..463F} and results from the Palomar Transient Factory (PTF), which remedy the bias against small galaxies by discovering the SNe first and the galaxy second, suggest systematic differences in CC SNe in small and normal galaxies. For example, more CC SNe have been discovered in dwarf galaxies than predicted in \cite{2008A&A...489..359Y} from the SFR and a CC SN mass range of $8$--$50 \, {\rm M_\odot}$ \citep{2010ApJ...721..777A}. Also, the distributions of CC SNe types are different \citep{2010ApJ...721..777A}. 

It is instructive to compare the distributions of CC SNe and SFR as functions of host galaxy properties. The fraction of CC SNe in dwarfs (defined as galaxies with $M_R > -18$ mag in the PTF) is $15/69 \approx 0.22$ \cite[Figure 3 of][]{2010ApJ...721..777A}, even though the share of the total SFR in such galaxies is only $0.1$ \citep[Figure 15 of][]{2004MNRAS.351.1151B}. For a cut of $M_R > -19$ mag, the fractions are $0.33$ and $0.20$, respectively. $K$ corrections are insufficient to account for this difference, and the difference could be interpreted as an increased efficiency of forming CC SNe in small galaxies. The increase is a factor of $\sim 2$ over that in normal galaxies, and the extra CC SNe would increase current SNR measurements. Since $20$\% of the SFR (in small galaxies) yields twice as many CC SNe, the increase in the SNR measurements is approximately $20$\%. As an extreme, we adopt an increase of $2 \times 0.2/0.8 \approx 50$\%, i.e., assuming that all CC SNe in small galaxies are unaccounted for at present. A detailed comparison of the distributions of CC SNe and SFR requires more data and is beyond the scope of this paper. The importance of small galaxies will become clearer with time, as SN surveys continue to discover more CC SNe in small galaxies. In particular, if the efficiency is enhanced over a wider range of galaxies, the increase in the SNR could be larger than discussed here. 

Finally, in all SN surveys, a short cadence is desirable for discovering SN during the most luminous epochs, as well as for obtaining a well sampled light curve for efficiency calculation purposes. The cadences of surveys, shown in Table \ref{table:SNRate}, are generally shorter than the peak durations of CC SNe. For example, Type Ibc typically drop by a magnitude in less than 1 month, while Type IIP evolve more slowly, with a plateau remaining within 1 mag of the peak for about 2--3 months. 

In summary, incomplete surveying of small galaxies may lead to SNR measurements underestimating the true SNR. Based on current data on dwarf galaxies, the effect may be a $20$--$50$\% increase in the SNR measurements. The importance of small galaxies remains to be investigated further with more data, including its caveats (Section \ref{sec:alternates}).

%======================================================================================
\subsection{Are CC SNe Dust Corrections Insufficient?}\label{sec:dust}

Host dust extinction makes CC SNe appear dimmer and correcting for missing obscured CC SNe is often the most uncertain ingredient in SNR measurements. Here we review dust corrections performed in SNR measurements. 

The dust correction applied to SNR measurements ranges from a few tens of percent at low redshifts to a factor of $\sim 2$ at high redshifts. According to the semi-analytic model of \cite{1998ApJ...502..177H}, the distribution of CC SN extinction due to host galaxy inclination peaks at low extinction but has an extended tail reaching many magnitudes ($19.8$ mag for a perfectly edge-on galaxy). This model was applied in SNLS, who found that it increased their measured SNR by $\approx 15$\% \citep{2009A&A...499..653B}. \cite{2008A&A...479...49B} used a statistical approach similar to \cite{1998ApJ...502..177H} and found an increase of a factor of $\sim 2$. Although the authors also present an extreme dust scheme which would in fact find agreement with the SNR predicted from the cosmic SFR, they point out that such an extreme dust scenario is not strongly motivated and is meant as an upper limit only. At higher redshift, \cite{2004ApJ...613..189D} combined the model of \cite{1998ApJ...502..177H} with a starburst extinction law, reporting increases of $60$\% and a factor of $\sim 2$ for their two measurements at $z \approx 0.3$ and $z \approx 0.7$, respectively. 

As pointed out in \cite{2007MNRAS.377.1229M}, at high redshift there is additional extinction due to starburst galaxies and highly star-forming galaxies (luminous and Ultraluminous IR galaxies, which we collectively call ULIRG). Identified by their strong FIR emission and high SFR, ULIRG should house many more CC SNe per galaxy, but we only detect a small fraction of CC SNe because of the higher dust obscuration. It has been estimated that as many as $60$\%--$90$\% of the CC SNe could go undiscovered \citep{2003A&A...401..519M}, although multi-wavelength studies are starting to discover some of these \citep{2008ApJ...689L..97K}. Because the importance of ULIRG increases with redshift, the fraction of CC SNe that are expected to be missing increases from $\sim 5$\% at $z \approx 0$ to $20$\%--$40$\% at $z \approx 1$ \citep{2007MNRAS.377.1229M}. 

In summary, correction for host galaxy obscuration remains uncertain. Extreme corrections that would explain the supernova rate problem are not prohibited, in particular at high redshifts where the excess extinction due to starburst galaxies is considered. However, this does not easily explain the supernova rate problem at low redshifts, where starburst galaxies provide only a small fraction of the SFR. We discuss missing dim CC SNe in more detail next. 

%======================================================================================
\subsection{Is the Contribution from Dim CC SNe Significant?}\label{sec:dim}

Type IIP SNe are the most common type of CC SNe and also the most varied in terms of luminosity \citep[e.g.,][]{2002AJ....123..745R,2003ApJ...582..905H}. The luminosity of a typical Type IIP SN remains nearly constant for a relatively long duration of $\sim 100$ days (the plateau phase), after which it drops sharply, marking the transition to the nebular phase \citep[e.g.,][]{1994A&A...282..731P}. It is expected that the radius of the progenitor plays a key role in shaping both the plateau length and luminosity \citep{1980ApJ...237..541A}. For example, the unusually dim light curve of SN 1987A is attributed to its compact progenitor \citep{1987ApJ...319..136A}. Recently, SN searches have revealed a class of Type IIP SNe that have peak and tail luminosities that are even lower than those of SN 1987A \citep{2003MNRAS.338..711Z,2004MNRAS.347...74P}. The prototype of these $^{56}$Ni-poor ($\lesssim 0.01 \, {\rm M_\odot}$) low-velocity Type IIP SNe is SN 1997D \citep[$M_V \approx -14.7$ mag;][]{1998ApJ...498L.129T}, and there are now more than a dozen similar CC SNe \citep[see, e.g.,][]{2004MNRAS.347...74P,2006MNRAS.370.1752P,2010arXiv1011.6558F}. The dimmest are SN 1999br \citep[$M_R \approx -13.5$ mag;][]{2004MNRAS.347...74P} and PTF10vdl/SN~2010id \citep[$M_R \approx -14.0$ mag;][]{2011ApJ...submit...G}, although even dimmer suspects exist, e.g., the transient in M85 ($M_R \approx -12$ mag; \citealp{2007Natur.449E...1P}, see also \citealp{2007Natur.447..458K} for an alternative scenario). Additionally, CC SNe dimmer than $M_R \approx -13.5$ mag have been inferred from the non-observation of SNe associated with several low-redshift long-GRBs \citep{2006Natur.444.1047F,2006Natur.444.1050D}. There is a population of dim ($M_V \lesssim -12$ mag) Type IIn SNe, but these are suspected to be LBV outbursts and not core collapses \citep[e.g.,][]{2006MNRAS.369..390M,2010arXiv1010.3718S}. 

In addition to the cadence and galaxy completeness effects mentioned in Section \ref{sec:luminous}, SN surveys are biased against low-luminosity CC SNe like SN~1997D. The limiting absolute magnitudes of SN surveys are typically $M \approx -15$ to $-16$ (Table \ref{table:SNRate}). Although SN surveys can discover dimmer CC SNe, the rapidly falling detection efficiency and the smaller volume from which dim CC SNe (we define CC SNe that are dimmer than $M = -15$ mag as ``dim'' CC SNe) may be confidently discovered makes collecting dim CC SNe challenging. To correct for missing dim CC SNe, a SN luminosity function that is complete to dimmer CC SNe must be adopted. Here, we revisit the importance of dim CC SNe. We make use of SN discoveries recorded in SN catalogs. The SN catalog combines results from SN surveys and amateur discoveries, resulting in a non-uniform sample of CC SNe of various qualities, bands and uncertainties. However, it is illustrative because it sets a strong lower bound on dim CC SNe. We first discuss the local ($\lesssim 100$ Mpc) volume for a qualitative analysis in Section \ref{sec:100Mpc}. Then, we focus on the very local ($\lesssim 10$ Mpc) volume for a more quantitative analysis in Section \ref{sec:10Mpc}.

%------------------------------------------------------------------------------------------------------------------------------------------------------
\subsubsection{How Ubiquitous are Dim CC SNe?}\label{sec:100Mpc}

We start with the Sternberg Astronomical Institute (SAI) SN Catalog \citep{2007HiA....14..316B}, selecting CC SNe of Types IIP, IIL, IIn, IIb, and Ibc. We select only the most recent decade, from 2000 to 2009, since previous decades have severe incompleteness problems \citep{2010ApJ...723..329H}. We cross check the classification with the Harvard SN database\footnote{http://www.cfa.harvard.edu/iau/lists/Supernovae.html}, which results in tagging some unclassified SNe as CC SNe. We update magnitudes to peak magnitudes where possible by investigating the literature and circulars\footnote{http://www.cfa.harvard.edu/iau/cbat.html}. We update SN host galaxy distances with redshift-independent distance measures from the Extragalactic Distance Database \citep[EDD;][]{2009AJ....138..323T}. When these are not available, we used distance estimates from NED\footnote{http://nedwww.ipac.caltech.edu/}. The uncertainties in the distances, which can reach $\pm 10$\%--$20$\%, affect the absolute magnitudes. As we discuss in Section \ref{sec:comparison}, this does not strongly affect our results.

%---------------------------------------------------------------
\begin{figure}[tb]
\centering\includegraphics[width=\linewidth,clip=true]{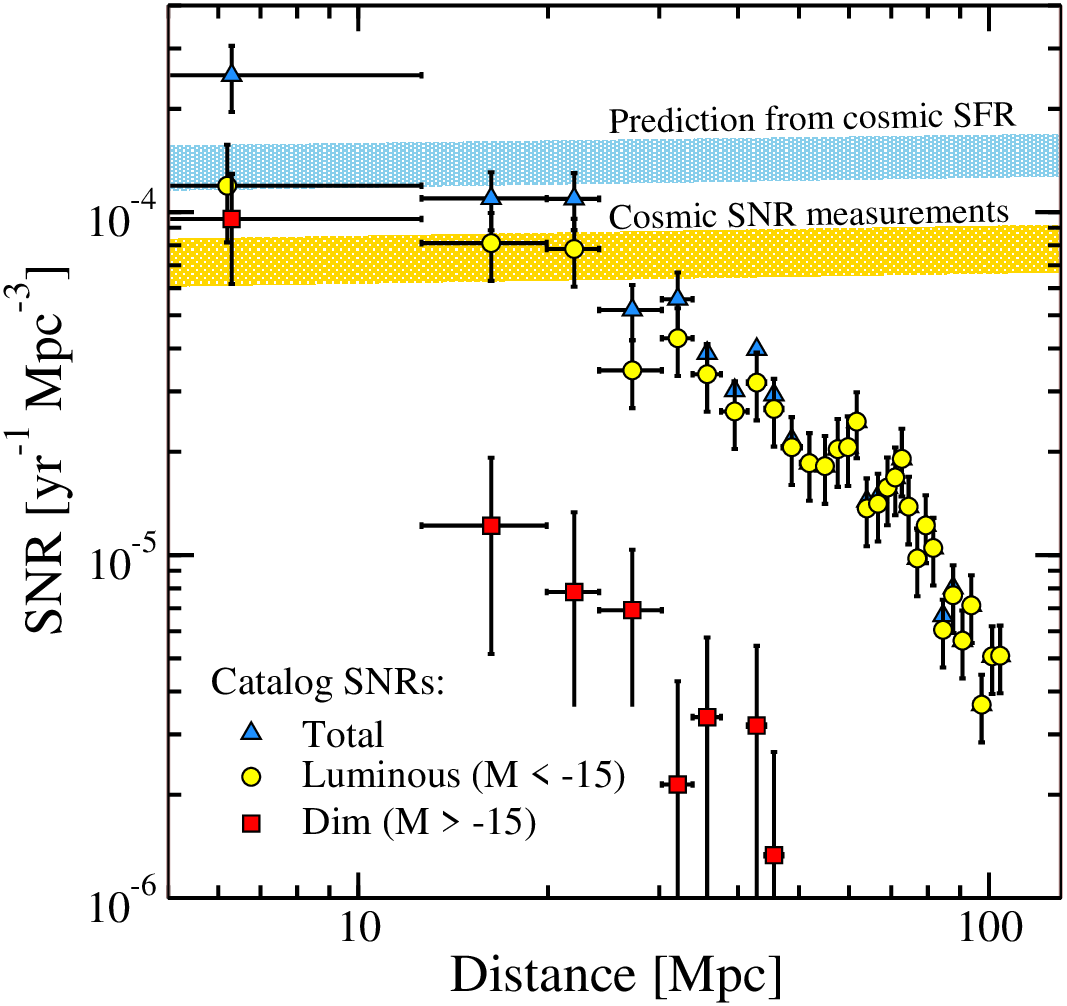}
\caption{Differential catalog SNR density as a function of distance. The total, luminous, and dim catalog SNR are shown by points (lowest distance bin has been shifted slightly in distance for clarity). The total catalog SNR is larger than the sum of the luminous and dim catalog SNR (see the text). Vertical error bars are statistical only. The SNR predicted from the cosmic SFR and directly measured are shown and labeled. The total catalog SNR is comparable to the upper band out to $\sim 25$ Mpc. The luminous catalog SNR is comparable to the lower band out to $\sim 25$ Mpc. The dim catalog SNR is large out to $\sim 10$ Mpc, indicating a significant fraction of dim CC SNe, independent of the absolute SFR enhancement within $\sim 10$ Mpc.} \label{fig:SNRate}
\end{figure}
%---------------------------------------------------------------

From the SAI catalog we discard CC SNe that seem to clearly be LBV transients (e.g., SN 2000ch, SN 2001ac, SN 2002kg, SN 2003gm, SN 2006fp, SN 2007sv, SN 2009ip). On the other hand, we include the possible SN impostors SN~2008S, NGC300-OT, and SN~2002bu, whose nature as either true CC SNe or extreme stellar outbursts remains subject to debate; we discuss the alternative assumption in Section \ref{sec:alternates}. Next, we check the discovery phases of CC SNe dimmer than $-16$ mag to identify CC SNe that are dim due to being discovered post-peak. For this identification we adopt the following criteria: SN IIP discovered more than 2 months post-peak, SN IIL over 1.5 months, and all SN IIn, IIb, Ibc over 1 month. These are the typical time scales over which the nominal light curves remain within 1 mag of the peak  \citep{2011MNRAS.412.1441L}. We identify 36 CC SNe that fit this category during 2000--2009 within $100$ Mpc.

From the modified SAI sample we construct three sets of ``catalog'' SNR. The first is the definitely luminous catalog SNR. For this we select CC SNe with absolute magnitudes below $-15$ mag. The second set is the definitely dim catalog SNR. For this we select CC SNe above $-15$ mag. We do not include dim CC SNe that have been discovered in their late phases, as the true peak magnitude is uncertain. The third and final set is the total catalog SNR. This includes all CC SNe, including definitely luminous CC SNe, definitely dim CC SNe, and late-time discoveries. Thus, the total catalog SNR is larger than the sum of the definitely luminous and definitely dim catalog SNR. 

\begin{deluxetable*}{llllllll}
\tablecaption{Very Local CC SNe During 2000 to 2009 Inclusive \label{table:local}} 
\tablehead{ 
SN		& Galaxy		& Type				& $D$ (Mpc)	& $E(B-V)$& Absolute Magnitude\tablenotemark{a}	& Discovery Phase	& References } 
\startdata
\hline
SN~2002bu	& NGC4242	& IIn\tablenotemark{b}	& 5.8		& 0.012	& $\bf M_R\approx-14.1$	& Early		& \cite{2002IAUC.7923....5H,2010arXiv1010.3718S}  \\  
SN~2002hh	& NGC6946	& IIP   				& 5.9		& 0.342	& $\bf M_R\approx-14.3$	& Early		& \cite{2006MNRAS.368.1169P} \\   
SN~2002kg	& NGC2403	& LBV				& 3.2		& 0.04	& $M_V\approx -9$		& Not CC SN	& \cite{2006MNRAS.369..390M} \\  
SN~2004am	& NGC3034 (M82)	& IIP 					& 3.5		& 0.159	& $M_R<-12.2$		& 3 months	& \cite{2004IAUC.8297....2S} \\ 
SN~2004dj	& NGC2403	& IIP    				& 3.2		& 0.04	& $M_R\sim-16.0$		& 1 month		& \cite{2006MNRAS.369.1780V,2006AJ....131.2245Z}  \\ 
SN~2004et	& NGC6946	& IIP					& 5.9		& 0.342	& $M_R\approx-17.6$	& Early		& \cite{2006MNRAS.372.1315S} \\ 
SN~2005af	& NGC4945	& IIP					& 3.6		& 0.177	& $\bf M_R\sim-15.4$	& 1 month		& \cite{2005IAUC.8482....1J} \\ 
SN~2005at	& NGC6744	& Ic					& 7.1		& 0.043	& $\bf M_R\sim-15.1$	& 2 weeks		& \cite{2005IAUC.8496....1M}  \\  
SN~2008bk	& NGC7793	& IIP  				& 4.1		& 0.019	& $\bf M_R\sim-15.5$	& 1 month		& \cite{2008CBET.1315....1M,2008CBET.1335....1M} \\  
SN~2008iz 	& NGC3034 (M82)	& II?   				& 3.5		& 0.159	& \bf{no optical}		& Radio only	& Brunthaler et al.~(2010)  \\  
SN~2008S	& NGC6946	& IIn\tablenotemark{b}	& 5.9		& 0.342	& $\bf M_R\approx-13.3$	& Early		& \cite{2009MNRAS.398.1041B,2009ApJ...697L..49S}   \\
NGC300-OT	& NGC300	& IIn\tablenotemark{b}	& 1.9		& 0.013	& $\bf M_V \sim -12.3$	& 1 month		& \cite{2009ApJ...695L.154B}  \\  
 \hline
SN~2002ap	& NGC0628	& IcPec				& 9.0		& 0.07	& $M_R \approx -17.8$	& Early		& \cite{2002MNRAS.332L..73G}  \\ 
SN~2003gd	& NGC0628	& IIP      				& 9.0		& 0.07	& $M_R \sim -16.7$		& 2 months	& \cite{2003IAUC.8152....3M} \\  
SN~2005cs	& NGC5194 (M51)	& IIP      				& 8.4		& 0.035	& $\bf M_R\approx-15.4$	& 1 month		& \cite{2006MNRAS.370.1752P}	\\  
SN~2007gr	& NGC1058	& Ic  	 				& 9.9		& 0.062	& $M_R \approx -17.4$	& Early		& \cite{2008ApJ...673L.155V} \\ 
SN~2008ax	& NGC4490	& IIb					& 9.6		& 0.022	& $M_R \approx -16.6$	& 2 weeks		& \cite{2008MNRAS.389..955P}  \\  
SN~2009hd	& NGC3627 (M66)	& IIP				& 8.3		& 0.032	& $\bf M_R\approx-13.9$	& Early		& \cite{2009CBET.1867....1M} \\
 \hline 
SN~2001ig	& NGC7424	& IIb	   				& 11.5	& 0.011	& $M_R \approx -17.3$	& Early		& \cite{2002IAUC.7804....2B}  \\  
SN~2003ie	& NGC4051	& II	   				& 12.2	& 0.013	& $M_R<-15.6$		& Uncertain	& \cite{2003IAUC.8205....1A} \\  
SN~2003jg	& NGC2997	& Ibc	   				& 11.3	& 0.109	& $\bf M_R\sim-14.1$	& Few weeks	& \cite{2003IAUC.8236....3B}  \\  
SN~2007it	& NGC5530	& IIP					& 11.7	& 0.116	& $M_V\approx-18.7$	& early		& \cite{2007IAUC.8875....1P}  \\  
SN~2008eh	& NGC2997	& Ibc?				& 11.3	& 0.109	& $\bf M_R \sim-15.3$	& 1 month		& \cite{2008CBET.1445....1M}  \\
SN~2009ib	& NGC1559	& IIP					& 12.6 	& 0.03	&$\bf M_R\approx-15.9$	& Early		& \cite{2009CBET.1902....1G}  \\ 
\tableline
\tablenotetext{a}{Shows the peak magnitudes for pre-peak discoveries, and the discovery or inferred peak magnitude for others. Values have been corrected for Galactic extinction. Unfiltered images are categorized as $R$-band.}
\tablenotetext{b}{Possible SN impostors, which may be CC SNe or stellar outbursts.}
\tablecomments{CC SNe that are definitely dimmer than $-16$ mag are shown in bold type. CC SNe with magnitude limits may be dim or luminous CC SNe, but cannot be convincingly identified with the data available. The list has been divided into blocks by distance ($7.9$, $10$, and $12.6$ Mpc). The first block and second block contain equal volumes, while the third block contains twice that.}
\end{deluxetable*}

All catalog SNR should be considered lower limits since they are derived from simple counting of likely incomplete SN discoveries. In addition to increasing incompleteness in older data \citep{2010ApJ...723..329H}, there is a strong observational bias against CC SNe in the southern hemisphere. In the latest decade and within $100$ Mpc, there are almost $1.9$ times more CC SNe discovered in the Northern hemisphere than the Southern hemisphere. There is also a strong bias against CC SNe in small galaxies. Considerations of the observational efficiency and coverage will only increase the catalog SNR. 

If the SN discoveries are sufficiently complete, the catalog SNR should be flat with distance. This is because they are number densities, and because the local SFR did not reveal significant cosmic variance, except in the very local volume within $\sim 10$ Mpc (Section \ref{sec:localSFR}). In other words, any significant decrease of the catalog SNR at any distance signifies an incompleteness of the SN discoveries at that distance. 

In Figure \ref{fig:SNRate}, the catalog SNR are binned in distance and compared to the cosmic SNR predicted from the SFR and directly measured. Neither host nor Galactic extinction corrections have been applied to the catalog SNR. Correcting would shift a small number of dim CC SNe to luminous CC SNe, but would not change qualitative conclusions; in Section \ref{sec:10Mpc}, where we discuss quantitative results, we apply Galactic extinction corrections. Each bin contains 20 luminous CC SNe, except for the smallest-distance bin, which contains 10 luminous CC SNe. 

The luminous catalog SNR should be compared to the measured cosmic SNR. The luminous catalog SNR is generally decreasing with distance but it is flat out to $\sim 25$ Mpc, suggesting reasonable completeness to that distance. The normalization of the flat section is comparable to the measured cosmic SNR on larger scales by LOSS, supporting this interpretation. There are 50 luminous CC SNe in the first 3 distance bins. The smallest-distance bin shows an enhancement as expected from the SFR.

The total catalog SNR should be compared to the predicted cosmic SNR. The total catalog SNR is flat out to $\sim 25$ Mpc and shows a normalization that is only slightly lower than expectations from the cosmic SFR. This supports our earlier claims that the true cosmic SNR is as large as expected \citep[e.g.,][]{2009PhRvD..79h3013H,2010ARNPS..60..439B}. There are $75$ CC SNe within the first three distance bins. Again, the smallest-distance bin shows an enhancement with respect to the predicted cosmic SNR, as expected from the SFR.

Finally, the dim catalog SNR falls with distance at all distances, a sign of incompleteness. At the smallest-distance bin, where dim CC SNe are least likely to be missed, the dim catalog SNR is, surprisingly, just as large as the luminous SNR measured by LOSS. There are $4$ dim CC SNe in this bin. This normalization is large enough to help solve the supernova rate problem, at least at $z \approx 0$, although the enhancement of the SFR in this distance range needs to be taken into account. 

Catalog CC SNe and Type Ia supernovae (SNe Ia) can be used to make an independent test of the normalization of the cosmic CC SNe rate \citep{2009PhRvD..79h3013H}. SNe Ia are luminous and have been intensively sought. Their cosmic rate history is better measured than that of CC SNe, and the distance out to which their catalog is apparently complete is larger \citep{2010ApJ...723..329H}. We consider the SN Ia to CC SN ratio, comparing rates from cosmic surveys near $z \approx 0$ (for larger $z$, the effect of SN Ia progenitor delays changes the ratio) and SN catalogs. From our fits to the measured cosmic rates, we find a SN Ia to CC SNe ratio at $z \approx 0$ of $\sim 0.18$. For comparison, the same ratio for the LOSS results is $\sim 0.43$. This difference of a factor of $\sim 2$ in the ratios is another statement of the supernova rate problem. The same ratio for the SN catalog should largely cancel incompleteness or local enhancement effects and is {\it independent of the assumed cosmic SFR}. We use a maximum distance large enough to contain enough SNe Ia and small enough that the CC SNe are reasonably complete. For SNe within $20$ Mpc, we find a ratio of $0.25$; this increases to $0.30$ by $25$ Mpc and rises to $0.35$ at larger distances. We interpret this as confirming that cosmic CC SNe are as common as predicted and that many dim ones are missed at increasing distances in catalogs and cosmic surveys. 

In summary, we analyze the local ($\lesssim 100$ Mpc) volume to study dim CC SNe. The SNR follows the trend expected from the SFR out to $\sim 25$ Mpc. We find that dim CC SNe are severely incomplete. Despite this, dim CC SNe are numerous in the very local ($\lesssim 10$ Mpc) volume. In the next Section, we discuss the fraction of dim CC SNe in more detail. 

%------------------------------------------------------------------------------------------------------------------------------------------------------
\subsubsection{What is the Fraction of Dim CC SNe within $\sim 10$ Mpc?}\label{sec:10Mpc}

We use the very local ($\lesssim 10$ Mpc) volume to quantify the importance of dim CC SNe. It is not feasible to extend to greater distances because of the severe incompleteness of dim CC SNe. We choose a nominal distance of $10$ Mpc, which yields sufficient statistics, and discuss how halving and doubling the volume changes the results. As discussed in the previous section, the normalization of the dim catalog SNR within $10$ Mpc is high. However, the very local volume has an enhanced absolute SFR. Therefore, we conservatively use the fraction of dim CC SNe, $f_{\rm dim}=N_{\rm dim}/(N_{\rm dim}+N_{\rm luminous})$, the ratio of definitely dim CC SNe over the sum of definitely dim and definitely luminous CC SNe. Here, the subscript $\rm dim$ refers to objects fainter than $\rm dim$-th magnitude. We will focus on a dim-luminous magnitude cut of $-15$ mag, but also show $-16$ mag for illustration. 

\begin{deluxetable}{lrrr}
\tablecaption{Summary of Very Local CC SNe \label{table:local2}} 
\tablehead{ 
Number of CC SNe					&  $0$--$7.9$ Mpc		& $7.9$--$10$ Mpc	& $10$--$12.6$ Mpc } 
\startdata
Expected total (cosmic rate)			& $2.9 \pm 0.5$		& $2.9 \pm 0.5$	& $5.8 \pm 1$	\\
Expected (LOSS rate)				& $1.5 \pm 0.4$		& $1.5 \pm 0.4$	& $3.0 \pm 0.9$  \\
Total observed CC SNe				& 12\tablenotemark{a}	& 6				& 6		\\
\hspace{12.3mm} $M < -16 $			& 2					& 4				& 2		\\
\hspace{3mm} $-16 < M < -15 $		& 3					& 1				& 2		\\
\hspace{3mm} $-15 < M $ (dim)		& 2					& 1				& 1		\\
\hspace{3mm} Possible SN impostor	& 3					& 0				& 0		\\
\hspace{3mm} Poor-quality CC SNe		& 2					& 0				& 1		\\
\tableline
\tablenotetext{a}{9 CC SNe if possible SN impostors are not included.}
\tablecomments{Expected number of CC SNe are shown for comparison. They are derived from measurements at larger distances, and should be multiplied by $\lesssim 2$ within $\sim 10$ Mpc to reflect the overdensity of SFR at these distances. The uncertainty is from the 1$\sigma$ uncertainty on the cosmic SFR or the uncertainty on the SNR measurements. }
\end{deluxetable}

The very local CC SNe are shown in Table \ref{table:local}. Here, the magnitudes are corrected for Galactic extinction following \cite{1998ApJ...500..525S} and assuming a \cite{1989ApJ...345..245C} Galactic reddening law. Distance estimates for CC SNe host galaxies vary among sources, and we adopt the estimates used in dedicated papers on the CC SNe when this is available. Otherwise, we continue to use the distances from EDD or NED as explained in Section \ref{sec:dim}. 

Several CC SNe are dim due to being highly extinguished by host dust, e.g., SN~2002hh with $A_V \approx 5 $ mag and SN 2008ax with $A_R\sim1$ mag \citep{2006MNRAS.368.1169P,2004IAUC.8299....2M}. The extreme case is SN~2008iz, which was discovered in radio observations and is not seen in the optical, suggesting an explosion behind a large gas or dust cloud in the central part of the host galaxy \citep{2010A&A...516A..27B}. We assume SN~2008eh to have been a CC SN even though it lacks a spectroscopic classification. Despite its relatively early discovery (it was not observed in the field 1 month prior to discovery), it was dim \citep[$M_R \sim -15.3$;][]{2008CBET.1445....1M}. From archival Spitzer images of its host galaxy NGC 2997, we locate the SN position in a spiral arm, next to a star forming region. Its light curve most closely resembles a dim Type Ibc CC SNe rather than a SN~2002cx-like peculiar Type Ia. 

It is not feasible to estimate the peak magnitudes of some CC SNe. For example, SN~2004am was discovered in the nebular phase, and the discovery phase of SN~2003ie is unknown. We collectively treat these CC SNe as poor-quality CC SNe. We exclude these CC SNe from estimates of $f_{\rm dim}$, since we want to estimate the dim fraction from a clean sample of SNe. 

Within $10$ Mpc, there are 18 CC SNe including 1 LBV (SN~2002kg). Excluding poor-quality CC SNe leaves 16 CC SNe, of which 6 are more luminous than $-16$ mag (SN~2002ap, SN~2003gd, SN~2004dj, SN~2004et, SN~2007gr, SN~2008ax) and 10 are dimmer than $-16$ mag. Of the latter, 3 are highly reddened (SN~2002hh, SN~2008iz, SN~2009hd), 4 are intrinsically dim (SN~2005af, SN~2005at, SN~2005cs, SN~2008bk), and 3 are possible SN impostors (SN~2002bu, SN~2008S, NGC300-OT). The number of CC SNe dimmer than $-15$ mag is $6$, so the dim fraction is $f_{>-15}=6/16$ ($f_{>-16}=10/16$). If we include poor-quality CC SNe and allow them to be either luminous or dim CC SNe, the dim fraction varies between $f_{>-15}=6/18$ and $f_{>-15}=8/18$.

%---------------------------------------------------------------
\begin{figure}[t]
\centering\includegraphics[width=\linewidth,clip=true]{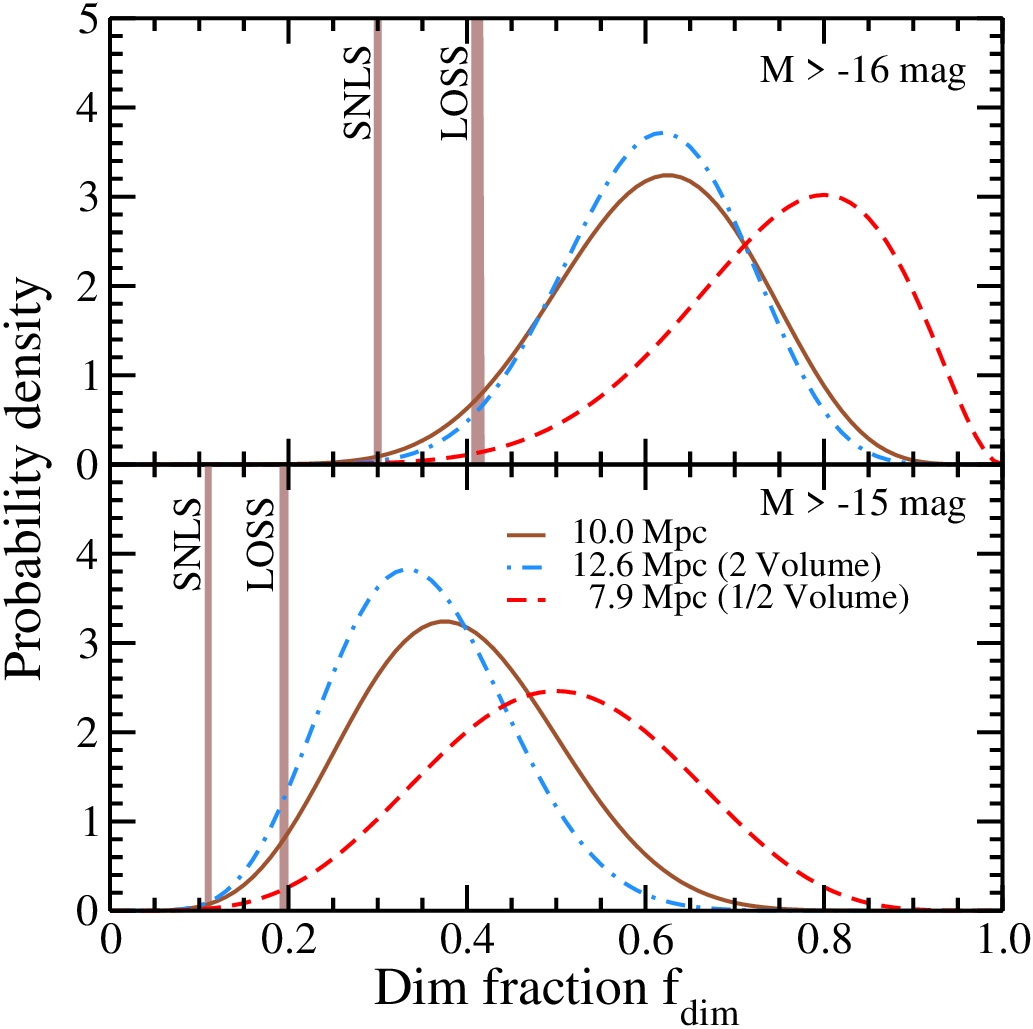}
\caption{Probability density distribution for the dim fraction ($f_{\rm dim}$), given the observed number of total and dim CC SNe, for three volume cuts. Shown for CC SNe with poor-quality CC SNe conservatively removed; see the text and Table \ref{table:local2}. The dim fraction is significantly higher than previous estimates from larger distances (vertical bands). Top: for CC SNe dimmer than $-16$ mag. Bottom: for those dimmer than $-15$ mag. } \label{fig:fraction}
\end{figure}
%---------------------------------------------------------------

We expect the dim fraction to decrease with distance, since dim CC SNe are more easily missed at larger distances. Going out to twice the volume (i.e., within $12.6$ Mpc) we find 6 more CC SNe (5 after excluding poor-quality CC SNe). There is 1 dim CC SNe (SN~2003jg), bringing the dim fraction to $f_{>-15}=7/21$ ($f_{>-16}=13/21$). Considering half the volume (i.e., CC SNe within $7.9$ Mpc) we find 12 CC SNe (10 after excluding poor-quality CC SNe), with $f_{>-15}=5/10$ ($f_{>-16}=8/10$). The CC SNe tallies are summarized in Table \ref{table:local2}. 

In Figure \ref{fig:fraction}, we plot the probability density function of $f_{\rm dim}$ given the observed number of total and dim CC SNe, for different distance cuts. We use the binomial distribution in $N_{\rm dim}$ for a fixed $N_{\rm tot}$ and varying $f_{\rm dim}$, using a flat prior in $f_{\rm dim}$ over the interval $0$--$1$. On the top panel, we show a magnitude cut of $-16$ mag for illustration. Due to the broad distribution of $f_{\rm dim}$, we are not able to derive a precise value for $f_{\rm dim}$ with the present data: the fractions are $f_{>-15} \sim 30$\%--$50$\%  ($f_{>-16} \sim 60$\%--$80$\%), but lower and higher values are not improbable. However, our estimates are clearly larger than previous studies based on more systematic samples at higher distances which were not as sensitive to the dim end of the CC SNe luminosity function ($f_{>-15} \sim 3$\%--$20$\%; Section \ref{sec:comparison}). Our estimates decrease to $f_{>-15} \sim 20$\%--$40$\% ($f_{>-16} \sim 50$\%--$70$\%) if possible SN impostors are excluded from our list of CC SN. This is still larger than previous studies. 

In summary, the very local $10$ Mpc volume shows the dim fraction could be as high as $f_{>-15} \sim 50$\%. This is significantly higher than previous estimates, which range between $f_{>-15} \sim 3$\%--$20$\%. Thus the SNR measurements are scaled up by $(1-0.2)/(1-0.5) = 1.6$ (a factor of $1.9$ if the original dim fraction is $5$\%), largely solving the supernova rate problem. While the very local $10$ Mpc volume is special in its absolute SFR, this should not affect the fraction of dim CC SNe. 

%%%%%%%%%%%%%%%%%%%%%%%%%%%%%%%%%%%%%%%%%%%%%%%%%%%%%%%%
%%%%%%%%%%%%%%%%%%%%%%%%%%%%%%%%%%%%%%%%%%%%%%%%%%%%%%%%
\section{Discussion}\label{sec:discussion}

In Figure \ref{fig:summary} we show the supernova rate problem at $z\approx0$. The bands reflect the nominal uncertainties on each quantity: for the predicted SNR it is the $1 \sigma$ uncertainty in the cosmic SFR fit by \cite{2006ApJ...651..142H}, while for the measured SNR it is the combined statistical and systematic uncertainty of the fit to SNR measurements. Although the supernova rate problem remains over the entire redshift range where SNR measurements are available, we focus here on the $z \approx 0$ range where there is most data. 

%======================================================================================
\subsection{Summary of Potential Explanations}

%---------------------------------------------------------------
\begin{figure}[tb]
\centering\includegraphics[width=\linewidth,clip=true]{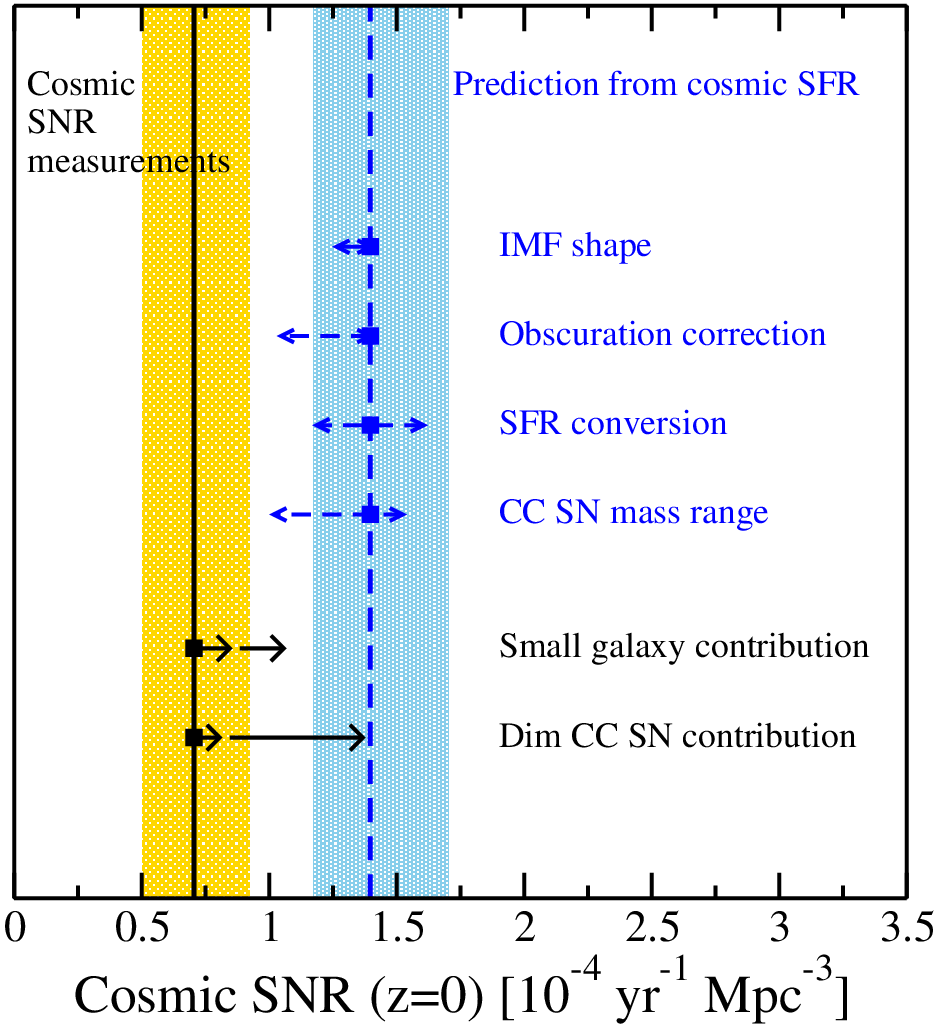}
\caption{Measured and predicted SNR at $z \approx 0$. Shaded bands reflect uncertainties: statistical and systematic combined for the measured SNR, and the $1\sigma$ uncertainty in the SFR for the predicted SNR. The results of varying the inputs for the predicted SNR are shown by dashed arrows, while the possible effects of missing CC SNe are shown for the measured SNR by solid arrows. Stacked arrows reflect different assumptions (see the text). The contribution from dim CC SNe can potentially increase the measured SNR to bridge the normalization discrepancy, although other effects can also be important.} \label{fig:summary}
\end{figure}
%---------------------------------------------------------------

First, to investigate whether the predicted SNR is too large, we explore the intrinsic uncertainties in the SFR measurements (IMF, obscuration, and conversion factor) and the additional factors when the SFR is converted to a SNR (IMF and CC SN mass range). The results are shown in Figure \ref{fig:summary} by dashed arrows.
\begin{itemize}

\item \emph{IMF}. We vary the IMF shape from the Salpeter A IMF to a classic Salpeter IMF and a flatter Baldry-Galzebrook IMF. This only has a small effect on the predicted SNR. 
\item \emph{Obscuration}. Obscuration correction is redshift dependent. Our adopted SFR compilation is corrected for obscuration by adding the FIR-derived SFR to the UV-derived SFR. We consider uncertainty due to two potential FIR contaminants: first, from AGN, which has been estimated to be $\lesssim 10$\% of the FIR \citep{2004MNRAS.355..973S}, and second, from cirrus emission, which we assume to be $\lesssim 50$\% of the FIR \citep{1987ApJ...314..513L}. We thus assume a reduction to the predicted SNR of $25$\%. Extinction uncertainties are unlikely to reduce the SNR prediction normalization by a full factor of $2$ (see Section \ref{sec:sfr} for integral constraints on the low redshift SFR normalization).
\item \emph{SFR conversion factor}. We assume formal uncertainties of $\pm 15$\%. Although conversion uncertainties for young star forming galaxies are larger, such galaxies are rare in the low-$z$ universe \citep{1991AJ....101..354S} and can be safely neglected.
\item \emph{CC SN mass range}. The nominal mass range is $8$--$40 \, \rm M_\odot$.  We use a minimum mass range taken from observations of CC SN progenitors: $8.5$--$16.5 \, \rm M_\odot$ and $25$--$40 \, \rm M_\odot$. We stress that this range does not account for the progenitors of all CC SNe Types. Further SN observations are needed to clarify this situation. We adopt the maximum mass range of $8$--$100 \, \rm M_\odot$.
\end{itemize}
Therefore, the uncertainties affecting the predicted SNR are generally smaller than the normalization discrepancy. 
 
Second, to investigate whether the measured SNR is too small, we explore the contribution from missing CC SNe (incomplete galaxy sample and dim CC SNe). This is shown by solid arrows in Figure \ref{fig:summary}.
\begin{itemize}
\item \emph{Incomplete galaxy sample}. This only affects SN surveys targeting a pre-selected sample of galaxies. Though careful work is done to correct for the bias against CC SNe in small galaxies, recent results suggest CC SNe in dwarf galaxies are unlike those in normal galaxies. We adopt a tentative factor of two enhancement in the CC SNe formation efficiency in dwarfs, which would yield a $20$\%--$50$\% increase in the total SNR, as shown by stacked arrows. Detailed studies need to await more data. 
\item \emph{Dim CC SNe}. There is an observational bias against the discovery and correction of dim CC SNe. In the very local ($\lesssim 10$ Mpc) volume, where dim CC SNe are least likely to be missed, the fraction of dim CC SNe is higher than in SN surveys targeting larger distances. Using the years 2000--2009, we find $f_{>-15} \sim 30$--$50$\% (if possible SN impostors are excluded from our list of CC SNe, we find $ 20$\%--$40$\%), whereas SN surveys find $3$\%--$20$\%. Therefore, we scale up the SNR measurement accordingly by a factor of between $(1-0.2)/(1-0.3) \approx 1.14$ and $(1-0.03)/(1-0.5) \approx 1.9$, shown by stacked arrows.
\end{itemize}
Therefore, if possible SN impostors are true CC SNe, the fraction of dim CC SNe can be sufficiently high that dim CC SNe largely explain the supernova rate problem. If possible SN impostors are not true CC SNe, $f_{\rm dim}$ is smaller, and dim CC SNe do not fully bridge the normalization discrepancy. 

In Figure \ref{fig:summary} we show how the different effects compare. The supernova rate problem is likely explained by a combination of the effects shown. 

%======================================================================================
\subsection{How Representative is the Very Local Volume?}\label{sec:representative}

The validity of using the dim fraction derived from very local ($\lesssim 10$ Mpc) CC SNe relies on how representative the very local volume is. We discuss several issues. 

First, distance uncertainties for the very local galaxies can be up to $\pm 25$\% in some cases. This has two effects: $(1)$ the absolute magnitude is affected and can cause erroneous categorization of CC SNe near the dim-luminous magnitude cut and $(2)$ CC SNe can be categorized in the wrong distance bin. We have considered the dim fraction using different magnitude cuts and different distance cuts, showing that it is consistently higher than previous estimates. 

Second, the very local volume shows an enhancement in its SNR (Figure \ref{fig:SNRate} and Table \ref{table:local2}). The enhancement is significant: within $12.6$ Mpc we observe $14$ luminous CC SNe while the expectation is $6$, an $\sim 0.2$\% occurrence. The enhancement is a factor of $\sim 2$, which is comparable to the factor of $1.7$ enhancement observed in the SFR in similar distances. The enhancement is observed not only in the luminous CC SNe but also the total CC SNe: there are $23$ discovered (including $3$ possible SN impostors) while the expectation is $12$. This can be interpreted as dim CC SNe are not severely incomplete within these distances.

Finally, to apply the dim fraction at cosmological distance requires some care. The IMF at high redshift remains poorly constrained \citep{2005MNRAS.363L..31N,2006MNRAS.365..712L}, and studies find that IMF evolution may be required at $z > 0.5$ \citep[e.g.,][]{2008MNRAS.391..363W}. If dim CC SNe are mass-dependent, an evolving IMF will lead to an evolving $f_{\rm dim}$. Also, redshift dependence may arise from metallicity dependence. A more important issue is obscuration by dust, which is known to be redshift-dependent (Section \ref{sec:dust}). Our work in the very local volume is mainly on normal galaxies, whose importance is overtaken by ULIRG at high redshifts \citep{2007MNRAS.377.1229M}. The higher dust obscuration in ULIRG causes as many as $90$\% of the CC SNe to be missed by current SN surveys \citep{2003A&A...401..519M}. Assuming that the dim fraction in ULIRG is $90$\%, the weighted dim fraction increases from $\sim50$\% at $z \approx 0$ to $\sim70$\% at $z\approx1$. 

In summary, we find that the very local ($\lesssim 10$ Mpc) volume has a high absolute SNR, which is related to an enhanced SFR. The dim fraction should be independent of the absolute normalization, and thus our estimate may be used at low redshifts. However, the dim fraction may increase with redshift.

%======================================================================================
\subsection{Dim CC SNe: Are They Compatible with Other Studies?} \label{sec:comparison}

One of the earliest discussions of dim CC SNe was based on historical CC SNe in the Milky Way and nearby galaxies \citep{1996ApJ...464..404S}. It was suggested that the CC SN luminosity function rises toward dimmer objects, with perhaps a third or more of the CC SNe dimmer than $M_B = -15$. However, the completeness of historical SNe has been difficult to assess, and the SN sample contains significant extinction and type uncertainties. Some of the dim CC SNe are now known to have been LBV, e.g., SN~1954J, for which the surviving star has been observed \citep{2005PASP..117..553V}. 

In more recent studies, the fractions are smaller. \cite{2002AJ....123..745R} report the mean and standard deviation of Type IIP, Type IIL, and Type Ibc SNe luminosity functions to be $M_B \approx -16.18$ ($\sigma \approx 1.23$), $M_B \approx -17.49$ ($\sigma \approx 1.29$), and $M_B \approx -17.37$ ($\sigma \approx 0.88$) (adjusted to our chosen cosmology). Using the type fractions of the volumetric SN sample of LOSS, the fraction of CC SNe dimmer than $M_B = -15$ ($-16$) is only $12$\% ($34$\%). In \cite{2008A&A...479...49B}, the authors adopt somewhat narrower luminosity functions which result in a smaller dim fraction of $3$\% ($13$\%). The SNLS uses an absolute magnitude scale relative to the mean magnitude of Type Ia SNe \citep{2009A&A...499..653B}. Adopting $M_V =-19$ mag as the mean Type Ia SN peak magnitude, the fraction of CC SNe dimmer than $M_V = -15$ ($-16$) is $11$\% ($30$\%). Here we use their luminosity function that has been corrected for the smaller target volume for dim CC SNe (data points of their Figure 12). The recently published luminosity functions of LOSS contain large numbers of dim CC SNe, including several in Table \ref{table:local}, e.g., SN~2002hh, SN~2004am, SN~2007gr, and SN~2008ax. From their luminosity functions, the fraction of CC SNe dimmer than $M_R = -15$ ($-16$) is $20$\% ($40$\%). Here, we have summed their luminosity histograms, rather than integrate their Gaussian fits which the authors caution are poor fits to the data \citep{2011MNRAS.412.1473L}. In summary, the dim fraction in the literature is $f_{>-15} \sim 3$\%--$20$\% ($f_{>-16} \sim 10$\%--$40$\%). In Figure \ref{fig:fraction} we label the fractions according to SNLS and LOSS, the two most recent and reliable measurements. 

Using the very local SN catalog data, we estimate the dim fraction to be $f_{>-15} \sim 30$\%--$50$\% ($f_{>-16} \sim 60$\%--$80$\%), larger than those of previous studies. Furthermore, we have assessed the completeness of the catalog CC SNe (Section \ref{sec:100Mpc}). The likelihood of our high $f_{\rm dim}$ being a simple statistical fluctuation is small given the sample size, e.g., the likelihood of fluctuating $f_{>-15} \sim 3$\% ($20$\%) to $f_{>-15} \gtrsim 50$\% in a sample size of $10$ CC SNe is $0.0005$\% ($3$\%). If possible SN impostors are excluded, $f_{>-15} \sim 20$\%--$40$\% ($f_{>-16} \sim 50$\%--$70$\%). This is in better agreement although still somewhat larger than the most recent luminosity function of LOSS. It is still significantly larger compared to other luminosity functions.

An important reason is the small distance cut adopted. In SN surveys targeting larger distances, the dimmest CC SNe are only discovered in a fraction of the volume corresponding to the smallest distance. Furthermore, a good understanding of the detection efficiency is required to correct the small statistics of dim CC SNe. For example, \cite{2002AJ....123..745R} suggests that the fraction of CC SNe dimmer than $M_B = -15$ is likely larger than $20$\% even though their nominal luminosity function suggests a value closer to $10$\%, because of the bias against dim SNe and small number statistics. On the other hand, we limit ourselves to a nominal distance of $10$ Mpc and the past decade from 2000 to 2009 to study dim CC SNe. The majority of SN searches in this very local volume have limiting magnitudes of 18 mag, corresponding to $\approx -12$ mag. Therefore, our very local sample is only limited by cadence and incompleteness of the galaxies observed. This may also explain why the dim fraction is lower at larger distances: by $30$--$40$ Mpc, the limiting magnitudes of most searches are already more luminous than $-15$ mag. Finally, we add that even so, our results should be treated as lower limits, since many of the CC SNe have been discovered by amateurs with different systematics, galaxy samples, and cadences, i.e., not systematically collected.

In summary, our estimate of $f_{\rm dim}$ is significantly larger than previous studies. If possible SN impostors are not true CC SNe, we estimate a value for $f_{\rm dim}$ that is somewhat larger than the SN luminosity function of LOSS.

%======================================================================================
\subsection{The Nature of Possible SN Impostors and Alternate Solutions to the Supernova Rate Problem?}\label{sec:alternates}

The dimmest objects in Table \ref{table:local} (SN~2002bu, SN~2008S, and NGC300-OT) are a special kind of explosion. Their nature---whether an extremely dim CC SNe or an extremely luminous stellar outburst---remains debated. Here, we discuss implications for the supernova rate problem if they are stellar outbursts. 

There is currently not enough data to conclude the true nature of possible SN impostors. \cite{2009ApJ...697L..49S} interpret SN~2008S as a super-Eddington wind of a $\sim 20 {\rm M_\odot}$ star, based on among others the lack of a  $^{56}{\rm Co} \to$ $^{56}{\rm Fe}$ decay tail. On the other hand, \cite{2009MNRAS.398.1041B} identify in their pseudo-bolometric late-time light curve a decay slope consistent with $^{56}{\rm Co}$ decay, concluding an electron-capture CC SNe of a $\sim 9 {\rm M_\odot}$ star. Measuring the bolometric light curve is a difficult multi-wavelength task, and further observations are required to settle the case. The progenitor of SN~2008S has been identified on Spitzer pre-explosion images to be a highly dust-enshrouded $\sim 10 {\rm M_\odot}$ star \citep{2008ApJ...681L...9P,2010arXiv1007.0011P}. Such stars are extremely rare \citep{2010ApJ...715.1094K}, suggesting it may be a common but short-lived phase prior to some kind of explosion \citep{2009ApJ...705.1364T}. The situations are similar for NGC300-OT \citep[][]{2009ApJ...695L.154B,2009ApJ...699.1850B} and SN~2002bu \citep{2009ApJ...705.1364T,2011ApJ...732...63S}.

If possible SN impostors are not CC SNe, the dim fraction is only somewhat larger than the luminosity function of LOSS, and missing dim CC SNe provide only a partial explanation to the supernova rate problem. In this case, there are several interpretations of the supernova rate problem, each with their own tensions with other measurements. We detail some of them below. 

First, the supernova rate problem may be interpreted as a large number of undetected CC SNe. For example, the fraction of optically dark CC SNe could be high \citep[see, e.g., Figure 3 of][for how the optically-dark CC SNe fraction relates to the SFR and measured SNR]{2010PhRvD..81h3001L}. Dark CC SNe as the solution to the supernova rate problem is only barely allowed by neutrino considerations \citep{2010PhRvD..81h3001L} and may be implausible on a theoretical basis \citep[e.g.,][]{2011ApJ...730...70O}. An alternate possibility is that more CC SNe than expected are driven below survey sensitivities by heavy host dust attenuation. However, this would appear difficult since such a change in dust properties would also increase the derived SFR and hence the predicted SNR, thus not necessarily helping solve the supernova rate problem. As pointed out by \cite{2008A&A...479...49B}, the consistency of the extinction model used in SFR and SNR measurements need to be further examined.

A second possibility is that CC SNe in small galaxies play a significantly larger role than expected. As discussed in Section \ref{sec:luminous}, the efficiency of forming CC SNe in dwarf galaxies may be higher than in normal galaxies. If the enhancement exists in a wider range of small galaxies, the effect of missing small galaxies could be larger than we estimate. However, there is a potential caveat with the CC SNe in small galaxies for helping solve the supernova rate problem. It is convenient to re-express the supernova rate problem in terms of the efficiency, or $N_{\rm CCSN}/M$, the number of CC SNe per unit stellar mass formed. The measured cosmic SNR and the measured cosmic SFR imply $0.0041/{\rm M_\odot}$, whereas the nominal prediction is $0.0088/{\rm M_\odot}$ (see Section \ref{sec:SNRate}); the factor of $\sim 2$ difference is another statement of the supernova rate problem. If the efficiency in small galaxies is taken to be enhanced, the observed $N_{\rm CCSN}$ is corrected upwards, so that the observed $N_{\rm CCSN}/M$ is closer to nominal predictions. However, the same enhanced efficiency will also increase the $N_{\rm CCSN}/M$ used in predictions, unless the efficiency in normal galaxies is sub-nominal. This is not seen in the LOSS galaxies, as \cite{2011MNRAS.tmp..307M} find $N_{\rm CCSN}/M = 0.0059$--$0.010/{\rm M_\odot}$ for a subset of LOSS galaxies with SDSS spectroscopic SFR measurements. Correcting for systematic effects, their best estimate is $N_{\rm CC}/M = 0.010 \pm 0.002/{\rm M_\odot}$, close to the nominal value. Therefore, enhanced CC SNe formation efficiencies in small galaxies may not bridge the supernova rate normalization discrepancy. Whether this is the case remains to be investigated with more data. For example, the known aperture bias from the finite SDSS fiber diameter is most critical for low-redshift galaxies like those in the LOSS survey \citep{2005PASP..117..227K}. This would tend to underestimate the SFR in the LOSS galaxies, potentially leading to an artificially larger $N_{\rm CCSN}/M$. This may allow some parameter space for CC SNe in small galaxies to help bridge the cosmic SNR normalization mismatch. 

The final possibility is that the cosmic SFR is overestimated. In particular, the observed efficiency in the LOSS galaxies, $N_{\rm CCSN}/M = 0.010 \pm 0.002/{\rm M_\odot}$, implies that there is very little room for missing CC SNe in the LOSS galaxies. The supernova rate problem could then be interpreted as an overestimated cosmic SFR. However, this may not help solve the supernova rate problem. Say there was some new effect causing the cosmic SFR to be overestimated, for example contamination of SFR estimators or systematic offset in SFR calibrations. The same effect would also cause the SFR in the LOSS galaxies to be overestimated, because the same SFR estimators are used. Thus, while reducing the cosmic SFR will increase the observed efficiency closer to nominal predictions, the efficiency in the LOSS galaxies will also increase to almost twice nominal predictions ($\sim 0.02/{\rm M_\odot}$). In other words, the supernova rate problem has only shifted.

Additionally, the nominal uncertainties in the SFR input physics do not easily tolerate a factor of $2$ reduction in the cosmic SFR (Section \ref{sec:sfr}). The largest uncertainty in the measured SFR is due to dust correction. However, to reduce the cosmic SFR sufficiently requires removing most of the dust correction. Apart from the need to redefine our understanding of the FIR universe, demanding no dust would also reduce the SNR measurements. The other, and perhaps more intriguing, possibility, is the presence of previously unknown contaminations to the SFR indicators. However, several SFR indicators used must be systematically contaminated by the same degree, making this seem unlikely. In either case, the implications for affecting the SFR estimators would go beyond the supernova rate problem to a wide range of topics.

%======================================================================================
\section{Summary and Implications}\label{sec:summary}

We find that the SNR predicted from the cosmic SFR, using nominal assumptions about the occurrence efficiency of optically luminous CC SNe, is higher than the SNR measured by SN surveys. Fits to the data demonstrate a normalization discrepancy of a factor of $\sim 2$ at the $\sim 2 \sigma$ level: we term this the ``supernova rate problem.'' At high redshifts ($z \gtrsim 0.4$) the measurement uncertainties are large and the significance is lower. We systematically explore resolutions. First, we investigate the various inputs and assumptions that feed into the SNR predictions and show that their uncertainties do not tolerate a normalization reduction by a factor of $\sim 2$. We then explore dim CC SNe (dimmer than $-15$ mag) that would have been difficult to discover in SNR measurements. From a sample of very local ($\lesssim 10$ Mpc) CC SNe, we estimate that as many as $\sim 50$\% of CC SNe may be dim, which could largely explain the SNR discrepancy. Our analysis is enabled by the recent precise measurements of the cosmic SFR and the SNR. In particular, the extensive SN survey by LOSS greatly contributed to the discovery of local CC SNe.

A large dim fraction has been suggested before by \cite{1996ApJ...464..404S} based on historical CC SNe. Here, we analyze more modern data and assess the completeness of the CC SNe sample. We find that the dim fraction is significantly larger than current luminosity functions used in SFR measurements. However, the exact dim fraction is still uncertain for several reasons: (1) the nature of the dimmest CC SNe are ambiguous at present. They appear to be unlike previous classified explosions and, if they are not true CC SNe, the dim fraction is more modest, closer to $\sim 30$\%; (2) dim CC SNe have not been systematically searched for, so many may have been missed. Until a systematic survey addresses this, our estimates for dim CC SNe should be treated as lower limits; (3) based on the current statistics, the dim fraction follows a fairly broad probability distribution; (4) finally, CC SNe could be missed for other reasons. For example, a fraction of CC SNe associated with the prompt formation of black holes could be even dimmer and not observable by current optical, IR, or radio surveys. 

The supernova rate problem raises some interesting implications. We are left with three main possible outcomes.
\begin{itemize}

\item {\bf First}, the fraction of dim CC SNe is high ($f_{>-15} \sim 30$\%--$50$\%). This is likely if the possible SN impostors are true CC SNe. The supernova rate problem is solved, with the implication that half of stars with masses  $8$--$40 \, {\rm M_\odot }$ are producing dim CC SNe, either due to dust obscuration or being intrinsically weak. 

\item {\bf Second}, if possible SN impostors are not CC SNe, the dim fraction is $f_{>-15} \sim 20$--$40$\%, i.e., only slightly higher than the most recent SN luminosity function of LOSS \citep{2011MNRAS.412.1473L}. The majority of dim CC SNe are due to heavily obscured CC SNe. The supernova rate problem could be explained by a high fraction of optically dark CC SNe, although the required fraction is somewhat higher than theoretical expectations \citep[e.g.,][]{2011ApJ...730...70O}. 

\item {\bf Third}, the dim fraction is $f_{>-15} \sim 20$\%--$40$\%, and the supernova rate problem is explained by systematic changes in our understanding of star formation or CC SN formation. For example, the efficiency of forming CC SNe in small galaxies may be higher than in normal galaxies. However, this may not help solve the supernova rate problem because the predicted SNR is also increased. Another example is that SFR estimators may be systematically overestimating the true cosmic SFR, although any explanation must go beyond current understanding of SFR estimation. If this turns out to be the cause, the implications would go beyond the supernova rate problem to a wide range of topics. 
\end{itemize}

To resolve these outcomes, detailed studies of the very local ($\lesssim 10$ Mpc) SFR using various indicators are needed to set the expected normalization of all CC SNe including optically luminous, dim, and dark CC SNe. The increased coverage of the Southern hemisphere by e.g., the CHilean Automatic Supernova sEarch \citep[CHASE; ][]{2009AIPC.1111..551P} will aid by providing a more complete discovery list of SNe. To verify the dim fraction, we need to capture all collapsing massive stars: optically luminous, dim, and dark. The All-Sky Automated Survey for the Brightest Supernovae \citep[ASAS-SN; e.g.,][]{2010arXiv1008.4126K,2011ApJ...730...34S} will find CC SNe in a volume-limited sample of nearby galaxies ($\lesssim 30$~Mpc) that will contribute to a better census of relatively low-luminosity events. In fact, one previously missed CC SN has already been identified in ASAS and CRTS images \citep{Prieto}. The PTF will collect a large number of early-discovered CC SNe and allow a high-statistics study of CC SNe varieties and host properties \citep{2009PASP..121.1395L,2009JCAP...01..047L,2010ApJ...721..777A}. Already, dim CC SNe (potentially dimmer than SN 1999br and PTF10vdl/SN 2010dl) have been discovered by the PTF \citep{GalYam2011}. Another interesting possibility is to observe the \emph{disappearance} of massive stars as opposed to the \emph{appearance} of their explosions, which will reveal the dark collapse rate \citep{2008ApJ...684.1336K}. Progenitor studies will be important in connecting future SFR and CC SNe studies. 

It will ultimately become possible to probe all kinds of collapsing stars with the deployment of next-generation non-electromagnetic probes such as neutrinos and gravitational waves \citep{2004APh....21..201A}. All core collapses---whether optically luminous or dim or dark---make neutrinos in comparable numbers. Proposed megaton neutrino detectors would be sensitive to core collapses in the several Mpc volume, revealing the true collapse rate of massive stars \citep{2005PhRvL..95q1101A,2008arXiv0810.1959K}. The nature of possible SN impostors could also be tested with the observation or non-observation of a neutrino signal. 

The diffuse SN neutrino background \citep[DSNB; see, e.g.,][]{2010ARNPS..60..439B,2010arXiv1007.3252L} similarly probes all core collapses and will test the true cosmic collapse rate of massive stars. The Super-Kamiokande limit on the DSNB \citep{2003PhRvL..90f1101M} is very near theoretical predictions using a cosmic SFR similar to the one assumed in this paper \citep{2009PhRvD..79h3013H}, and prospects for detecting the DSNB will be significantly increased by the near-future Gadolinium-enhanced Super-Kamiokande \citep{2004PhRvL..93q1101B,2008ICRC....5.1421W}. We emphasize that while the luminous SNR in cosmic surveys is lower than expected, there is no evidence for a reduction in the predicted DSNB flux, which depends on all core collapses. If the neutrino emission from a collapse to a black hole is significantly higher in energy compared to the canonical collapse to a neutron star \citep[e.g., as discussed in][]{2008PhRvD..78h3014N}, the DSNB will be sensitive to the fraction of black hole forming collapses \citep[e.g.,][]{2010PhRvD..81h3001L,2010arXiv1012.1274K}. 

Using the full array of SN surveys, progenitor studies, survey of disappearing stars, DSNB searches, and upcoming gravitational wave detectors, a comprehensive understanding of the deaths of massive stars and their associated observable transient phenomena will come within reach in the coming years. This will be an important complement to studies of the progenitors of SN Ia and their explosion mechanisms that will be opened by similar optical SN studies together with next-generation MeV gamma-ray detectors \citep[e.g.,][]{2010ApJ...723..329H}.

%%%%%%%%%%%%%%%%%%%%%%%%%%%%%%%%%%%%%%%%%%%%%%%%%%%%%%%%
%%%%%%%%%%%%%%%%%%%%%%%%%%%%%%%%%%%%%%%%%%%%%%%%%%%%%%%%

\acknowledgments
We thank Berto Monard for sharing photometry of SN~2008eh. We thank Matt Bothwell, Tim Eifler, Alex Filippenko, Avishay Gal-Yam, Robert Kennicutt, Matt Kistler, Weidong Li, Amy Lien, Dan Maoz, Michael Mortenson, Jim Rich, and Stephen Smartt for discussions. This research made use of the IAU Central Bureau for Astronomical Telegrams and the Sternberg Astronomical Institute supernova catalogs and the NASA/IPAC Extragalactic Database (NED), which is operated by JPL/Caltech, under contract with NASA. 
SH and JFB were supported by NSF CAREER Grant PHY-0547102 (to JFB); CSK, KZS, and TAT by NSF Grant AST-0908816; and JLP by NASA through Hubble Fellowship Grant HF-51261.01-A awarded by STScI, which is operated by AURA, Inc. for NASA, under contract NAS~5-2655.

%%%%%%%%%%%%%%%%%%%%%%%%%%%%%%%%%%%%%%%%%%%%%%%%%%%%%%%%
%%%%%%%%%%%%%%%%%%%%%%%%%%%%%%%%%%%%%%%%%%%%%%%%%%%%%%%%

%%%%%%%%%%%%%%%%%%%%%%%%%%%%%%%%%%%%%%%%%%%%%%%%%%%%%%%%
%%%%%%%%%%%%%%%%%%%%%%%%%%%%%%%%%%%%%%%%%%%%%%%%%%%%%%%%
%\clearpage


\begin{thebibliography}{99}

\bibitem[Abdo et al.(2010)]{2010ApJ...723.1082A} Abdo, A.~A., et al.\ 2010, \apj, 723, 1082 

\bibitem[Aharonian et al.(2006)]{2006Natur.440.1018A} Aharonian, F., et al.\ 2006, \nat, 440, 1018 

\bibitem[Albert et al.(2008)]{2008Sci...320.5884A} Albert, J., et al.\ 2008, Science, 320, 5884 

\bibitem[Anderson \& James(2008)]{2008MNRAS.390.1527A} Anderson, J.~P., \& James, P.~A.\ 2008, \mnras, 390, 1527 

\bibitem[Anderson \& James(2009)]{2009MNRAS.399..559A} Anderson, J.~P., \& James, P.~A.\ 2009, \mnras, 399, 559 

\bibitem[Ando et al.(2005)]{2005PhRvL..95q1101A} Ando, S., Beacom, J.~F., Y\"{u}ksel, H.\ 2005, Physical Review Letters, 95, 171101 

\bibitem[Arbour \& Boles(2003)]{2003IAUC.8205....1A} Arbour, R., \& Boles, T.\ 2003, \iaucirc, 8205, 1 

\bibitem[Arcavi et al.(2010)]{2010ApJ...721..777A} Arcavi, I., et al.\ 2010, \apj, 721, 777 

\bibitem[Arnaud et al.(2004)]{2004APh....21..201A} Arnaud, N., et al.\ 2004, Astroparticle Physics, 21, 201 

\bibitem[Arnett(1987)]{1987ApJ...319..136A} Arnett, W.~D.\ 1987, \apj, 319, 136 

\bibitem[Arnett(1980)]{1980ApJ...237..541A} Arnett, W.~D.\ 1980, \apj, 237, 541 

\bibitem[Arnouts et al.(2005)]{2005ApJ...619L..43A} Arnouts, S., et al.\ 2005, \apjl, 619, L43 

\bibitem[Avni(1976)]{1976ApJ...210..642A} Avni, Y.\ 1976, \apj, 210, 642 

\bibitem[Baldry \& Glazebrook(2003)]{2003ApJ...593..258B} Baldry, I.~K., \& Glazebrook, K.\ 2003, \apj, 593, 258 

\bibitem[Baldry et al.(2005)]{2005MNRAS.358..441B} Baldry, I.~K., et al.\ 2005, \mnras, 358, 441 

\bibitem[Bartunov et al.(2007)]{2007HiA....14..316B} Bartunov, O.~S., Tsvetkov, D.~Y., \& Pavlyuk, N.~N.\ 2007, Highlights of Astronomy, 14, 316 

\bibitem[Bazin et al.(2009)]{2009A&A...499..653B} Bazin, G., et al.\ 2009, \aap, 499, 653 

\bibitem[Beacom \& Vagins(2004)]{2004PhRvL..93q1101B} Beacom, J.~F., \& Vagins, M.~R.\ 2004, Physical Review Letters, 93, 171101

\bibitem[Beacom(2010)]{2010ARNPS..60..439B} Beacom, J.~F.\ 2010, Annual Review of Nuclear and Particle Science, 60, 439 

\bibitem[Bell(2003)]{2003ApJ...586..794B} Bell, E.~F.\ 2003, \apj, 586, 794 

\bibitem[Bembrick et al.(2002)]{2002IAUC.7804....2B} Bembrick, C., Pearce, A., \& Evans, R.\ 2002, \iaucirc, 7804, 2 

\bibitem[Berger et al.(2009)]{2009ApJ...699.1850B} Berger, E., et al.\ 2009, \apj, 699, 1850

\bibitem[Biggs et al.(2003)]{2003IAUC.8236....3B} Biggs, J., Monard, L.~A.~G., \& Africa, S.\ 2003, \iaucirc, 8236, 3 

\bibitem[Bond et al.(2009)]{2009ApJ...695L.154B} Bond, H.~E., Bedin, L.~R., Bonanos, A.~Z., Humphreys, R.~M., Monard, L.~A.~G.~B., Prieto, J.~L., \& Walter, F.~M.\ 2009, \apjl, 695, L154 

\bibitem[Bothwell et al.(2011)]{2011arXiv1104.0929B} Bothwell, M.~S., et al.\ 2011, arXiv:1104.0929 

\bibitem[Botticella et al.(2008)]{2008A&A...479...49B} Botticella, M.~T., et al.\ 2008, \aap, 479, 49 

\bibitem[Botticella et al.(2009)]{2009MNRAS.398.1041B} Botticella, M.~T., et al.\ 2009, \mnras, 398, 1041 

\bibitem[Brinchmann et al.(2004)]{2004MNRAS.351.1151B} Brinchmann, J., Charlot, S., White, S.~D.~M., Tremonti, C., Kauffmann, G., Heckman, T., \& Brinkmann, J.\ 2004, \mnras, 351, 1151 

\bibitem[Brunthaler et al.(2010)]{2010A&A...516A..27B} Brunthaler, A., et al.\ 2010, \aap, 516, A27 

\bibitem[Cappellaro et al.(1999)]{1999A&A...351..459C} Cappellaro, E., Evans, R., \& Turatto, M.\ 1999, \aap, 351, 459 

\bibitem[Cappellaro et al.(2005)]{2005A&A...430...83C} Cappellaro, E., et al.\ 2005, \aap, 430, 83 

\bibitem[Cardelli et al.(1989)]{1989ApJ...345..245C} Cardelli, J.~A., Clayton, G.~C., \& Mathis, J.~S.\ 1989, \apj, 345, 245 

\bibitem[Condon et al.(2002)]{2002AJ....124..675C} Condon, J.~J., Cotton, W.~D., \& Broderick, J.~J.\ 2002, \aj, 124, 675 

\bibitem[Dahlen et al.(2004)]{2004ApJ...613..189D} Dahlen, T., et al.\ 2004, \apj, 613, 189 

\bibitem[Dahlen et al.(2010)]{2010AAS...21543023D} Dahlen, T., Strolger, L., \& Riess, A.~G.\ 2010, Bulletin of the American Astronomical Society, 42, 360 

\bibitem[Della Valle et al.(2006)]{2006Natur.444.1050D} Della Valle, M., et al.\ 2006, \nat, 444, 1050 

\bibitem[Elias-Rosa et al.(2010)]{2010ApJ...714L.254E} Elias-Rosa, N., et al.\ 2010, \apjl, 714, L254 

\bibitem[Essey \& Kusenko(2010)]{2010APh....33...81E} Essey, W., \& Kusenko, A.\ 2010, Astroparticle Physics, 33, 81 

\bibitem[Essey et al.(2011)]{2011ApJ...731...51E} Essey, W., Kalashev, O., Kusenko, A., \& Beacom, J.~F.\ 2011, \apj, 731, 51 

\bibitem[Fardal et al.(2007)]{2007MNRAS.379..985F} Fardal, M.~A., Katz, N., Weinberg, D.~H., \& Dav{\'e}, R.\ 2007, \mnras, 379, 985 

\bibitem[Ferrara \& Ricotti(2006)]{2006MNRAS.373..571F} Ferrara, A., \& Ricotti, M.\ 2006, \mnras, 373, 571 

\bibitem[Fioc \& Rocca-Volmerange(1997)]{1997A&A...326..950F} Fioc, M., \& Rocca-Volmerange, B.\ 1997, \aap, 326, 950 

\bibitem[Fraser et al.(2010a)]{2010ApJ...714L.280F} Fraser, M., et al.\ 2010a, \apjl, 714, L280 

\bibitem[Fraser et al.(2010b)]{2010arXiv1011.6558F} Fraser, M., et al.\ 2010b, arXiv:1011.6558 

\bibitem[Fruchter et al.(2006)]{2006Natur.441..463F} Fruchter, A.~S., et al.\ 2006, \nat, 441, 463 

\bibitem[Fryer(1999)]{1999ApJ...522..413F} Fryer, C.~L.\ 1999, \apj, 522, 413 

\bibitem[Fynbo et al.(2006)]{2006Natur.444.1047F} Fynbo, J.~P.~U., et al.\ 2006, \nat, 444, 1047 

\bibitem[Gal-Yam et al.(2002)]{2002MNRAS.332L..73G} Gal-Yam, A., Ofek, E.~O., \& Shemmer, O.\ 2002, \mnras, 332, L73 

\bibitem[Gal-Yam et al.(2007)]{2007ApJ...656..372G} Gal-Yam, A., et al.\ 2007, \apj, 656, 372 

\bibitem[Gal-Yam \& Leonard(2009)]{2009Natur.458..865G} Gal-Yam, A., \& Leonard, D.~C.\ 2009, \nat, 458, 865 

\bibitem[Gal-Yam et al.(2011)]{2011ApJ...submit...G} Gal-Yam, A., et al.\ 2011, \apj, submitted

\bibitem[Gal-Yam (2011)]{GalYam2011} Gal-Yam, A.\ 2011, private communication

\bibitem[Gallego et al.(1995)]{1995ApJ...455L...1G} Gallego, J., Zamorano, J., Aragon-Salamanca, A., \& Rego, M.\ 1995, \apjl, 455, L1 

\bibitem[Giavalisco et al.(2004)]{2004ApJ...600L.103G} Giavalisco, M., et al.\ 2004, \apjl, 600, L103 

\bibitem[Graur et al.(2011)]{2011arXiv1102.0005G} Graur, O., et al.\ 2011, arXiv:1102.0005 

\bibitem[Hamuy(2003)]{2003ApJ...582..905H} Hamuy, M.\ 2003, \apj, 582, 905 

\bibitem[Hanish et al.(2006)]{2006ApJ...649..150H} Hanish, D.~J., et al.\ 2006, \apj, 649, 150 

\bibitem[Hatano et al.(1998)]{1998ApJ...502..177H} Hatano, K., Branch, D., \& Deaton, J.\ 1998, \apj, 502, 177 

\bibitem[Hauser \& Dwek(2001)]{2001ARA&A..39..249H} Hauser, M.~G., \& Dwek, E.\ 2001, \araa, 39, 249 

\bibitem[Heger et al.(2003)]{2003ApJ...591..288H} Heger, A., Fryer, C.~L., Woosley, S.~E., Langer, N., \& Hartmann, D.~H.\ 2003, \apj, 591, 288 

\bibitem[Hopkins(2004)]{2004ApJ...615..209H} Hopkins, A.~M.\ 2004, \apj, 615, 209 

\bibitem[Hopkins \& Beacom(2006)]{2006ApJ...651..142H} Hopkins, A.~M., \& Beacom, J.~F.\ 2006, \apj, 651, 142

\bibitem[Hopkins et al.(2006)]{2006ApJS..163....1H} Hopkins, P.~F., Hernquist, L., Cox, T.~J., Di Matteo, T., Robertson, B., \& Springel, V.\ 2006, \apjs, 163, 1 

\bibitem[Horiuchi et al.(2009)]{2009PhRvD..79h3013H} Horiuchi, S., Beacom, J.~F., \& Dwek, E.\ 2009, \prd, 79, 083013 

\bibitem[Horiuchi \& Beacom(2010)]{2010ApJ...723..329H} Horiuchi, S., \& Beacom, J.~F.\ 2010, \apj, 723, 329 

\bibitem[Hornoch(2002)]{2002IAUC.7923....5H} Hornoch, K.\ 2002, \iaucirc, 7923, 5 

\bibitem[Iglesias-P{\'a}ramo et al.(2006)]{2006ApJS..164...38I} Iglesias-P{\'a}ramo, J., et al.\ 2006, \apjs, 164, 38 

\bibitem[Jacques \& Pimentel(2005)]{2005IAUC.8482....1J} Jacques, C., \& Pimentel, E.\ 2005, \iaucirc, 8482, 1 

\bibitem[James et al.(2008)]{2008A&A...482..507J} James, P.~A., Knapen, J.~H., Shane, N.~S., Baldry, I.~K., \& de Jong, R.~S.\ 2008, \aap, 482, 507 

\bibitem[Janka et al.(2007)]{2007PhR...442...38J} Janka, H.-T., Langanke, K., Marek, A., Mart{\'{\i}}nez-Pinedo, G., {\ Muuml}ller, B.\ 2007, \physrep, 442, 38 

\bibitem[Jarosik et al.(2011)]{2011ApJS..192...14J} Jarosik, N., et al.\ 2011, \apjs, 192, 14

\bibitem[Kalirai et al.(2008)]{2008ApJ...676..594K} Kalirai, J.~S., Hansen, B.~M.~S., Kelson, D.~D., Reitzel, D.~B., Rich, R.~M., \& Richer, H.~B.\ 2008, \apj, 676, 594 

\bibitem[Kankare et al.(2008)]{2008ApJ...689L..97K} Kankare, E., et al.\ 2008, \apjl, 689, L97 

\bibitem[Karachentsev et al.(2004)]{2004AJ....127.2031K} Karachentsev, I.~D., Karachentseva, V.~E., Huchtmeier, W.~K., \& Makarov, D.~I.\ 2004, \aj, 127, 2031 

\bibitem[Keehn \& Lunardini(2010)]{2010arXiv1012.1274K} Keehn, J.~G., \& Lunardini, C.\ 2010, arXiv:1012.1274 

\bibitem[Kennicutt(1998)]{1998ARA&A..36..189K} Kennicutt, R.~C., Jr.\ 1998, \araa, 36, 189 

\bibitem[Kennicutt et al.(2008)]{2008ApJS..178..247K} Kennicutt, R.~C., Jr., Lee, J.~C., Funes, S.~J., Jos{\'e} G., Sakai, S., \& Akiyama, S.\ 2008, \apjs, 178, 247 

\bibitem[Kewley et al.(2002)]{2002AJ....124.3135K} Kewley, L.~J., Geller, M.~J., Jansen, R.~A., \& Dopita, M.~A.\ 2002, \aj, 124, 3135 

\bibitem[Kewley et al.(2005)]{2005PASP..117..227K} Kewley, L.~J., Jansen, R.~A., \& Geller, M.~J.\ 2005, \pasp, 117, 227 

\bibitem[Khan et al.(2010)]{2010ApJ...715.1094K} Khan, R., Stanek, K.~Z., Prieto, J.~L., Kochanek, C.~S., Thompson, T.~A., \& Beacom, J.~F.\ 2010, \apj, 715, 1094 

\bibitem[Khan et al.(2011)]{2010arXiv1008.4126K} Khan, R., et al.\ 2011, \apj, 726, 106 

\bibitem[Kistler et al.(2008)]{2008arXiv0810.1959K} Kistler, M.~D., Yuksel, H., Ando, S., Beacom, J.~F., \& Suzuki, Y.\ 2008, arXiv:0810.1959 

\bibitem[Kochanek et al.(2008)]{2008ApJ...684.1336K} Kochanek, C.~S., Beacom, J.~F., Kistler, M.~D., Prieto, J.~L., Stanek, K.~Z., Thompson, T.~A., Y{\"u}ksel, H.\ 2008, \apj, 684, 1336 

\bibitem[Kochanek et al.(2010)]{2010arXiv1010.3704K} Kochanek, C.~S., Szczygiel, D.~M., \& Stanek, K.~Z.\ 2010, arXiv:1010.3704 

\bibitem[Kotake et al.(2006)]{2006RPPh...69..971K} Kotake, K., Sato, K., \& Takahashi, K.\ 2006, Reports on Progress in Physics, 69, 971

\bibitem[Kulkarni et al.(2007)]{2007Natur.447..458K} Kulkarni, S.~R., et al.\ 2007, \nat, 447, 458 

\bibitem[Lagache et al.(2005)]{2005ARA&A..43..727L} Lagache, G., Puget, J.-L., \& Dole, H.\ 2005, \araa, 43, 727 

\bibitem[Lattimer \& Prakash(2007)]{2007PhR...442..109L} Lattimer, J.~M., \& Prakash, M.\ 2007, \physrep, 442, 109

\bibitem[Law et al.(2009)]{2009PASP..121.1395L} Law, N.~M., et al.\ 2009, \pasp, 121, 1395 

\bibitem[Le Delliou et al.(2006)]{2006MNRAS.365..712L} Le Delliou, M., Lacey, C.~G., Baugh, C.~M., \& Morris, S.~L.\ 2006, \mnras, 365, 712 

\bibitem[Le Floc'h et al.(2003)]{2003A&A...400..499L} Le Floc'h, E., et al.\ 2003, \aap, 400, 499 

\bibitem[Le Floc'h et al.(2005)]{2005ApJ...632..169L} Le Floc'h, E., et al.\ 2005, \apj, 632, 169 

\bibitem[Leaman et al.(2011)]{2011MNRAS.412.1419L} Leaman, J., Li, W., Chornock, R., \& Filippenko, A.~V.\ 2011, \mnras, 412, 1419 

\bibitem[Li et al.(2011a)]{2011MNRAS.412.1441L} Li, W., et al.\ 2011, \mnras, 412, 1441 

\bibitem[Li et al.(2011b)]{2011MNRAS.412.1473L} Li, W., Chornock, R., Leaman, J., Filippenko, A.~V., Poznanski, D., Wang, X., Ganeshalingam, M., \& Mannucci, F.\ 2011, \mnras, 412, 1473 

\bibitem[Ly et al.(2011)]{2011ApJ...726..109L} Ly, C., Lee, J.~C., Dale, D.~A., Momcheva, I., Salim, S., Staudaher, S., Moore, C.~A., \& Finn, R.\ 2011, \apj, 726, 109 

\bibitem[Lien \& Fields(2009)]{2009JCAP...01..047L} Lien, A., \& Fields, B.~D.\ 2009, JCAP, 1, 47 

\bibitem[Lien et al.(2010)]{2010PhRvD..81h3001L} Lien, A., Fields, B.~D., \& Beacom, J.~F.\ 2010, \prd, 81, 083001 

\bibitem[Lonsdale Persson \& Helou(1987)]{1987ApJ...314..513L} Lonsdale Persson, C.~J., \& Helou, G.\ 1987, \apj, 314, 513 

\bibitem[Lunardini(2010)]{2010arXiv1007.3252L} Lunardini, C.\ 2010, arXiv:1007.3252 

\bibitem[Madau et al.(1998)]{1998ApJ...498..106M} Madau, P., Pozzetti, L., \& Dickinson, M.\ 1998, \apj, 498, 106 

\bibitem[Malek et al.(2003)]{2003PhRvL..90f1101M} Malek, M., et al.\ 2003, Physical Review Letters, 90, 061101 

\bibitem[Mannucci et al.(2003)]{2003A&A...401..519M} Mannucci, F., et al.\ 2003, \aap, 401, 519 

\bibitem[Mannucci et al.(2007)]{2007MNRAS.377.1229M} Mannucci, F., Della Valle, M., \& Panagia, N.\ 2007, \mnras, 377, 1229 

\bibitem[Maoz et al.(2011)]{2011MNRAS.tmp..307M} Maoz, D., Mannucci, F., Li, W., Filippenko, A.~V., Della Valle, M., \& Panagia, N.\ 2011, \mnras, 412, 1508 

\bibitem[Martin et al.(2005)]{2005ApJ...619L..59M} Martin, D.~C., et al.\ 2005, \apjl, 619, L59 

\bibitem[Martin et al.(2005b)]{2005IAUC.8496....1M} Martin, R., Yamaoka, H., Monard, L.~A.~G., \& Africa, S.\ 2005, \iaucirc, 8496, 1 

\bibitem[Mattila et al.(2004)]{2004IAUC.8299....2M} Mattila, S., Meikle, W.~P.~S., Groeningsson, P., Greimel, R., Schirmer, M., Acosta-Pulido, J.~A., \& Li, W.\ 2004, \iaucirc, 8299, 2 

\bibitem[Massey et al.(2000)]{2000AJ....119.2214M} Massey, P., Waterhouse, E., \& DeGioia-Eastwood, K.\ 2000, \aj, 119, 2214

\bibitem[Massey et al.(2001)]{2001AJ....121.1050M} Massey, P., DeGioia-Eastwood, K., \& Waterhouse, E.\ 2001, \aj, 121, 1050 

\bibitem[Matteucci \& Greggio(1986)]{1986A&A...154..279M} Matteucci, F., \& Greggio, L.\ 1986, \aap, 154, 279 

\bibitem[Maund et al.(2006)]{2006MNRAS.369..390M} Maund, J.~R., et al.\ 2006, \mnras, 369, 390 

\bibitem[McNaught(2003)]{2003IAUC.8152....3M} McNaught, R.~H.\ 2003, \iaucirc, 8152, 3 

\bibitem[Miyaji et al.(1980)]{1980PASJ...32..303M} Miyaji, S., Nomoto, K., Yokoi, K., \& Sugimoto, D.\ 1980, \pasj, 32, 303 

\bibitem[Monard(2008a)]{2008CBET.1315....1M} Monard, L.~A.~G.\ 2008a, Central Bureau Electronic Telegrams, 1315, 1 

\bibitem[Monard(2008b)]{2008CBET.1445....1M} Monard, L.~A.~G.\ 2008b, Central Bureau Electronic Telegrams, 1445, 1 

\bibitem[Monard(2009)]{2009CBET.1867....1M} Monard, L.~A.~G.\ 2009, Central Bureau Electronic Telegrams, 1867, 1 

\bibitem[Morrell \& Stritzinger(2008)]{2008CBET.1335....1M} Morrell, N., \& Stritzinger, M.\ 2008, Central Bureau Electronic Telegrams, 1335, 1

\bibitem[Nagamine et al.(2004)]{2004ApJ...610...45N} Nagamine, K., Cen, R., Hernquist, L., Ostriker, J.~P., \& Springel, V.\ 2004, \apj, 610, 45 

\bibitem[Nagashima et al.(2005)]{2005MNRAS.363L..31N} Nagashima, M., Lacey, C.~G., Okamoto, T., Baugh, C.~M., Frenk, C.~S., \& Cole, S.\ 2005, \mnras, 363, L31 

\bibitem[Nakazato et al.(2008)]{2008PhRvD..78h3014N} Nakazato, K., Sumiyoshi, K., Suzuki, H., \& Yamada, S.\ 2008, \prd, 78, 083014 

\bibitem[Nomoto(1984)]{1984ApJ...277..791N} Nomoto, K.\ 1984, \apj, 277, 791 

\bibitem[O'Connor \& Ott(2011)]{2011ApJ...730...70O} O'Connor, E., \& Ott, C.~D.\ 2011, \apj, 730, 70 

\bibitem[Orr et al.(2011)]{2011ApJ...733...77O} Orr, M.~R., Krennrich, F., \& Dwek, E.\ 2011, \apj, 733, 77 

\bibitem[Pascale et al.(2009)]{2009ApJ...707.1740P} Pascale, E., et al.\ 2009, \apj, 707, 1740 

\bibitem[Pastorello et al.(2004)]{2004MNRAS.347...74P} Pastorello, A., et al.\ 2004, \mnras, 347, 74 

\bibitem[Pastorello et al.(2006)]{2006MNRAS.370.1752P} Pastorello, A., et al.\ 2006, \mnras, 370, 1752 

\bibitem[Pastorello et al.(2007)]{2007Natur.449E...1P} Pastorello, A., et al.\ 2007, \nat, 449, 1

\bibitem[Pastorello et al.(2008)]{2008MNRAS.389..955P} Pastorello, A., et al.\ 2008, \mnras, 389, 955 

\bibitem[Patat et al.(1994)]{1994A&A...282..731P} Patat, F., Barbon, R., Cappellaro, E., \& Turatto, M.\ 1994, \aap, 282, 731

\bibitem[Peacock \& Dodds(1994)]{1994MNRAS.267.1020P} Peacock, J.~A., \& Dodds, S.~J.\ 1994, \mnras, 267, 1020 

\bibitem[P{\'e}rez-Gonz{\'a}lez et al.(2003)]{2003ApJ...591..827P} P{\'e}rez-Gonz{\'a}lez, P.~G., Zamorano, J., Gallego, J., Arag{\'o}n-Salamanca, A., \& Gil de Paz, A.\ 2003, \apj, 591, 827

\bibitem[Pignata(2008)]{2009CBET.1902....1G} Pignata, G.\ 2009, Central Bureau Electronic Telegrams, 1902, 1 

\bibitem[Pignata et al.(2009)]{2009AIPC.1111..551P} Pignata, G., et al.\ 2009, American Institute of Physics Conference Series, 1111, 551 

\bibitem[Prieto et al.(2008)]{2008ApJ...681L...9P} Prieto, J.~L., et al.\ 2008, \apjl, 681, L9 

\bibitem[Prieto et al.(2010)]{2010arXiv1007.0011P} Prieto, J.~L., Szczygiel, D.~M., Kochanek, C.~S., Stanek, K.~Z., Thompson, T.~A., Beacom, J.~F., Garnavich, P.~M., \& Woodward, C.~E.\ 2010, arXiv:1007.0011 

\bibitem[Prieto et al.(2011)]{Prieto} Prieto, J.~L., et al.\ 2011, in prep.

\bibitem[Poelarends et al.(2008)]{2008ApJ...675..614P} Poelarends, A.~J.~T., Herwig, F., Langer, N., \& Heger, A.\ 2008, \apj, 675, 614 

\bibitem[Pojmanski(2007)]{2007IAUC.8875....1P} Pojmanski, G.\ 2007, \iaucirc, 8875, 1 

\bibitem[Pozzo et al.(2006)]{2006MNRAS.368.1169P} Pozzo, M., et al.\ 2006, \mnras, 368, 1169 

\bibitem[Raffelt(1990)]{1990PhR...198....1R} Raffelt, G.~G.\ 1990, \physrep, 198, 1 

\bibitem[Raffelt(2000)]{2000PhR...333..593R} Raffelt, G.~G.\ 2000, \physrep, 333, 593

\bibitem[Rich (2010)]{Rich2011} Rich, J.\ 2011, private communication

\bibitem[Richardson et al.(2002)]{2002AJ....123..745R} Richardson, D., Branch, D., Casebeer, D., Millard, J., Thomas, R.~C., \& Baron, E.\ 2002, \aj, 123, 745 

\bibitem[Rujopakarn et al.(2010)]{2010ApJ...718.1171R} Rujopakarn, W., et al.\ 2010, \apj, 718, 1171 

\bibitem[Sahu et al.(2006)]{2006MNRAS.372.1315S} Sahu, D.~K., Anupama, G.~C., Srividya, S., \& Muneer, S.\ 2006, \mnras, 372, 1315 

\bibitem[Salim et al.(2007)]{2007ApJS..173..267S} Salim, S., et al.\ 2007, \apjs, 173, 267 

\bibitem[Schaefer(1996)]{1996ApJ...464..404S} Schaefer, B.~E.\ 1996, \apj, 464, 404 

\bibitem[Schiminovich et al.(2005)]{2005ApJ...619L..47S} Schiminovich, D., et al.\ 2005, \apjl, 619, L47 

\bibitem[Schiminovich et al.(2007)]{2007ApJS..173..315S} Schiminovich, D., et al.\ 2007, \apjs, 173, 315 

\bibitem[Schlegel et al.(1998)]{1998ApJ...500..525S} Schlegel, D.~J., Finkbeiner, D.~P., \& Davis, M.\ 1998, \apj, 500, 525

\bibitem[Serjeant et al.(2002)]{2002MNRAS.330..621S} Serjeant, S., Gruppioni, C., \& Oliver, S.\ 2002, \mnras, 330, 621 

\bibitem[Silva et al.(2004)]{2004MNRAS.355..973S} Silva, L., Maiolino, R., \& Granato, G.~L.\ 2004, \mnras, 355, 973 

\bibitem[Singer et al.(2004)]{2004IAUC.8297....2S} Singer, D., Pugh, H., \& Li, W.\ 2004, \iaucirc, 8297, 2 

\bibitem[Smartt(2009)]{2009ARA&A..47...63S} Smartt, S.~J.\ 2009, \araa, 47, 63 

\bibitem[Smartt et al.(2009)]{2009MNRAS.395.1409S} Smartt, S.~J., Eldridge, J.~J., Crockett, R.~M., \& Maund, J.~R.\ 2009, \mnras, 395, 1409 

\bibitem[Smith et al.(2009)]{2009ApJ...697L..49S} Smith, N., et al.\ 2009, \apjl, 697, L49 

\bibitem[Smith et al.(2010)]{2010arXiv1010.3718S} Smith, N., Li, W., Silverman, J.~M., Ganeshalingam, M., \& Filippenko, A.~V.\ 2010, arXiv:1010.3718 

\bibitem[Smith et al.(2011)]{2011ApJ...732...63S} Smith, N., et al.\ 2011, \apj, 732, 63 

\bibitem[Soifer \& Neugebauer(1991)]{1991AJ....101..354S} Soifer, B.~T., \& Neugebauer, G.\ 1991, \aj, 101, 354 

\bibitem[Somerville et al.(2001)]{2001MNRAS.320..504S} Somerville, R.~S., Primack, J.~R., \& Faber, S.~M.\ 2001, \mnras, 320, 504 

\bibitem[Stoll et al.(2011)]{2011ApJ...730...34S} Stoll, R., Prieto, J.~L., Stanek, K.~Z., Pogge, R.~W., Szczygie{\l}, D.~M., Pojma{\'n}ski, G., Antognini, J., \& Yan, H.\ 2011, \apj, 730, 34 

\bibitem[Taylor et al.(2009)]{2009AAS...21342502T} Taylor, M., Cinabro, D., \& SDSS-II Supernova Survey Team 2009, Bulletin of the American Astronomical Society, 41, 252 

\bibitem[Thompson et al.(2009)]{2009ApJ...705.1364T} Thompson, T.~A., Prieto, J.~L., Stanek, K.~Z., Kistler, M.~D., Beacom, J.~F., \& Kochanek, C.~S.\ 2009, \apj, 705, 1364 

\bibitem[Todini \& Ferrara(2001)]{2001MNRAS.325..726T} Todini, P., \& Ferrara, A.\ 2001, \mnras, 325, 726 

\bibitem[Tully et al.(2009)]{2009AJ....138..323T} Tully, R.~B., Rizzi, L., Shaya, E.~J., Courtois, H.~M., Makarov, D.~I., \& Jacobs, B.~A.\ 2009, \aj, 138, 323 

\bibitem[Turatto et al.(1998)]{1998ApJ...498L.129T} Turatto, M., et al.\ 1998, \apjl, 498, L129 

\bibitem[Valenti et al.(2008)]{2008ApJ...673L.155V} Valenti, S., et al.\ 2008, \apjl, 673, L155 

\bibitem[Van Dyk et al.(2005)]{2005PASP..117..553V} Van Dyk, S.~D., Filippenko, A.~V., Chornock, R., Li, W., \& Challis, P.~M.\ 2005, \pasp, 117, 553 

\bibitem[Vink{\'o} et al.(2006)]{2006MNRAS.369.1780V} Vink{\'o}, J., et al.\ 2006, \mnras, 369, 1780 

\bibitem[Watanabe(2008)]{2008ICRC....5.1421W} Watanabe, H.\ 2008, International Cosmic Ray Conference, 5, 1421 

\bibitem[Watson et al.(2009)]{2009ApJ...696.2206W} Watson, C.~R., et al.\ 2009, \apj, 696, 2206 

\bibitem[Wilkins et al.(2008a)]{2008MNRAS.391..363W} Wilkins, S.~M., Hopkins, A.~M., Trentham, N., \& Tojeiro, R.\ 2008, \mnras, 391, 363 

\bibitem[Wilkins et al.(2008b)]{2008MNRAS.385..687W} Wilkins, S.~M., Trentham, N., \& Hopkins, A.~M.\ 2008, \mnras, 385, 687 

\bibitem[Williams et al.(2009)]{2009ApJ...693..355W} Williams, K.~A., Bolte, M., \& Koester, D.\ 2009, \apj, 693, 355 

\bibitem[Wolf et al.(2003)]{2003A&A...401...73W} Wolf, C., Meisenheimer, K., Rix, H.-W., Borch, A., Dye, S., \& Kleinheinrich, M.\ 2003, \aap, 401, 73 

\bibitem[Woosley(1993)]{1993ApJ...405..273W} Woosley, S.~E.\ 1993, \apj, 405, 273 

\bibitem[Young et al.(2008)]{2008A&A...489..359Y} Young, D.~R., Smartt, S.~J., Mattila, S., Tanvir, N.~R., Bersier, D., Chambers, K.~C., Kaiser, N., \& Tonry, J.~L.\ 2008, \aap, 489, 359 

\bibitem[Y{\"u}ksel et al.(2008)]{2008ApJ...683L...5Y} Y{\"u}ksel, H., Kistler, M.~D., Beacom, J.~F., \& Hopkins, A.~M.\ 2008, \apjl, 683, L5 

\bibitem[Yun et al.(2001)]{2001ApJ...554..803Y} Yun, M.~S., Reddy, N.~A., \& Condon, J.~J.\ 2001, \apj, 554, 803 

\bibitem[Zampieri et al.(2003)]{2003MNRAS.338..711Z} Zampieri, L., Pastorello, A., Turatto, M., Cappellaro, E., Benetti, S., Altavilla, G., Mazzali, P., \& Hamuy, M.\ 2003, \mnras, 338, 711 

\bibitem[Zhang et al.(2006)]{2006AJ....131.2245Z} Zhang, T., Wang, X., Li, W., Zhou, X., Ma, J., Jiang, Z., \& Chen, J.\ 2006, \aj, 131, 2245 

\end{thebibliography}
\end{document}